\documentclass[conference]{IEEEtran}

\usepackage{cite}
\usepackage{graphicx}
\usepackage{dcolumn}
\usepackage{amsmath}
\usepackage{amssymb}
\usepackage{bm}
\usepackage{color}
\usepackage[absolute,showboxes]{textpos}
\newcommand{\cref}[1]{(\ref{#1})}
\newcommand{\ud}{\mathrm{d}}

\TPMargin{5pt}

\newcommand{\copyrightstatement}{
    \begin{textblock}{0.84}(0.08,0.93)    
         \noindent
         \footnotesize
         \copyright 2015 IEEE. Personal use of this material is permitted. Permission from IEEE must be obtained for all other uses, in any current or future media, including reprinting/republishing this material for advertising or promotional purposes, creating new collective works, for resale or redistribution to servers or lists, or reuse of any copyrighted component of this work in other works.
    \end{textblock}
}

\begin{document}

\copyrightstatement

\title{Diffusiophoretic Self-Propulsion for Partially Catalytic Spherical Colloids}

\author{\IEEEauthorblockN{Joost de Graaf}
\IEEEauthorblockA{Institute for Computational Physics (ICP)\\
University of Stuttgart\\ 
Allmandring 3, 70569 Stuttgart, Germany \\
Email: jgraaf@icp.uni-stuttgart.de}
\and
\IEEEauthorblockN{Georg Rempfer\\ and Christian Holm}
\IEEEauthorblockA{Institute for Computational Physics (ICP)\\
University of Stuttgart\\ 
Allmandring 3, 70569 Stuttgart, Germany}
}

\maketitle

\begin{abstract}
Colloidal spheres with a partial platinum surface coating perform auto-phoretic motion when suspended in hydrogen peroxide solution. We present a theoretical analysis of the self-propulsion velocity of these particles using a continuum multi-component, self-diffusiophoretic model. With this model as a basis, we show how the slip-layer approximation can be derived and in which limits it holds. First, we consider the differences between the full multi-component model and the slip-layer approximation. Then the slip model is used to demonstrate and explore the sensitive nature of the particle's velocity on the details of the molecule-surface interaction. We find a strong asymmetry in the dependence of the colloid's velocity as a function of the level of catalytic coating, when there is a different interaction between the solute and solvent molecules and the inert and catalytic part of the colloid, respectively. The direction of motion can even be reversed by varying the level of the catalytic coating. Finally, we investigate the robustness of these results with respect to variations in the reaction rate near the edge between the catalytic and inert parts of the particle. Our results are of significant interest to the interpretation of experimental results on the motion of self-propelled particles.
\end{abstract}

\section{\label{sec:intro}Introduction}

Over the past decade, research into self-propelled particles (SPPs) has undergone rapid development and attracted strong interest from the scientific community, due to the inherent out-of-equilibrium nature of their behavior.~\cite{Cates_12,Cates_14} Examples of naturally occurring self-propelled `particles' include humans,~\cite{Helbing_00,Zhang_13,Sliverberg_13} birds,~\cite{Ballerini_08} fish,~\cite{Katz_11} insects,~\cite{Buhl_06,John_04} spermatozoa,~\cite{Woolley_03,Riedel_05,Ma_14} bacteria,~\cite{Sokolov_07,Schwarz-Linek_12,Reufer_14} and algae.~\cite{Polin_09,Geyer_13} The principle of autonomous movement is thus relevant across many orders of magnitude in size and speed. Some of the most interesting behavior of SPPs is observed when the particles are colloidal in size ($\sim 1$ nm -- 1 $\mu$m) and suspended in a liquid medium, since there will be a competition between Brownian (thermal) motion and the self-propulsion mechanism.~\cite{Lee_14} Moreover, the viscous nature of the suspending medium strongly impacts the strategies by which autonomous movement may be achieved mechanically. For low-Reynolds-number swimmers this movement must be nonreciprocal, as put forward by Purcell in his scallop theorem.~\cite{purcell}

There exists another set of colloidal SPPs that differs strongly from the aforementioned microorganisms, namely, artificial particles that move by self-generated solute gradients. Artificial SPPs were pioneered by the work of Ismagilov~\emph{et al.}~\cite{ismagilov} and Paxton~\emph{et al.}~\cite{Paxton_04} These early studies have led to the development of a wide range of artificial SPPs that move by utilizing a variety of gradient-based propulsion mechanisms, including: bimetallic nanorods that move by self-electrophoresis,~\cite{Paxton_04,Wang_06} Pt-coated colloidal spheres moving by self-diffusio- or self-electrophoresis,~\cite{Brown,Ebbens,Ebbens_12,Howse_07,Valadares_10,simmchen} Au-coated colloidal spheres that move by thermophoresis~\cite{Yoshinaga,Baraban_13a} or by thermally induced local mixing-demixing transitions,~\cite{Buttinoni_12} hollow cones that expel oxygen bubbles,~\cite{Solovev_09,Mei_11} and many others. A common theme for most gradient-based SPPs is the decomposition of hydrogen peroxide at a platinum surface, which acts as a catalyst to the reaction. These particles are considered ideally suited as model systems for out-of-equilibrium phenomena~\cite{Cates_14} and may be utilized as functional components in micro-fluidic devices,~\cite{Perez_14} or be employed for medical purposes.~\cite{Wang_14} However, there are many open questions with regards to the way these particles achieve their motion~\cite{Brown,Ebbens} and how this motion may be most effectively controlled,~\cite{Baraban_12b,Zhao_12,islam_14} that must be answered to allow for the successful implementation of these particles in real-world applications.

From the modeling perspective, a lot of attention has gone into the description of the self-propulsion mechanism of these particles. In particular, it is accepted that self-electrophoresis plays a dominant role for the self-propulsion of bi-metallic nanorods.~\cite{Moran_10,Sabass_12b} The mechanism by which Pt-coated colloidal spheres achieve motion in hydrogen-peroxide solution is, however, not as well understood. This has led to heated debate on the type of phoretic motion by which these particles self-propel.~\cite{Brown,Ebbens} The traditional view is that self-diffusiophoresis can be used to explain the experimentally observed motion in these systems,~\cite{Howse_07,Ebbens_12} which has resulted in a large number of studies into the specifics of this mechanism.

The majority of the theoretical models are based on a coarse-grained, continuum-level approach that employs the dilute-limit slip-layer formalism, see, e.g., Refs.~\cite{Anderson,Brady_11}. From a simulation/computation perspective, particle-based models are more commonly used.~\cite{Roeckner_07,Tao_08,Valadares_10,Lugli_11,Yang_11,Yang_14} Investigations into the properties of the self-diffusiophoretic model have considered the influence of a large number of parameters. For instance, the influence of the reaction mechanism on the size-velocity dependence of the SPPs was matched to experiment.~\cite{Howse_07,Ebbens_12} In addition, there have been more fundamental studies which unified the continuum- and particle-based descriptions~\cite{Brady_11,Mood_13b} and considered the influence of advective and reactive processes on the self-propulsion.~\cite{Lauga_14} Moreover, the influence of particle shape~\cite{Popescu_10,Lugli_11,Brady_14} and catalytic coating~\cite{Popescu_11,Mood_13,Lauga_14} have been considered, as well as the nature of the interaction potential.~\cite{Sabass_12a,Mood_13} 

A common theme in continuum modeling is that typically only the `effective' interaction between (one of) the solutes and the surface of the particle is taken into account for the slip-based self-propulsion. In addition, the interaction between the solute and the inert, as well as the solute and the catalytic surface area is often assumed the same. There are theoretical studies that have touched upon such differences, see, e.g., Refs.~\cite{Golestanian_07,Uspal_14,Mood_13}, but a full analysis is still lacking. The use of an identical interaction potential for the inert and catalytic surface in theory stands in stark contrast with the more common particle-based simulation models, wherein anisotropy of the solvent-surface interaction is often a crucial ingredient to achieve movement.~\cite{Roeckner_07,Tao_08,Valadares_10,Lugli_11,Yang_14} Furthermore, from a physical perspective, the assumption of a homogeneous molecule-surface interaction is questionable.

In this manuscript, we solve a set of multi-component diffusion-advection equations~\cite{Curtiss} with appropriate boundary conditions~\cite{Anderson,Brady_11} to describe the self-diffusiophoretic movement of a self-propelled particle (SPP). We depart from the traditional single-component dilute-limit approximation in order to maintain momentum conservation and incompressibility of the total fluid at high fuel (hydrogen-peroxide) and product (water and oxygen) concentrations. We derive the continuity equations governing the dynamics of the individual species from a chemical potential and show the approximations that lead to a self-consistent model. The role of the solvent-surface potential is elucidated and it is shown how the slip-layer approximation follows from these equations. In addition, we consider the differences with previously established multi-component models. Our advection-diffusion-reaction model is subsequently used to analyze the motion of a catalytically coated colloid suspended in hydrogen-peroxide solution. In particular, we investigate the sensitivity of the self-propulsion velocity on the surface-molecule interaction potential and the level of catalytic coating for physically reasonable parameters. This part of the investigation is in a similar vein as the work of Sabass~\emph{et al.}~\cite{Sabass_12a}, but takes a different approach to quantifying the effect of the surface-molecule interaction by coupling it to the level of catalytic coating. 

We find a strong sensitivity of the SPP's velocity on the molecule-surface interaction by taking into account the gradients of all the species involved in the reaction and by assigning different interaction potentials between the molecules and the inert and catalytic side of the particle, respectively. In contrast to the result of Refs.~\cite{Popescu_10,Mood_13}, we show that the typical dependence of the swimming velocity on the level of catalytic surface coating is asymmetric in this parameter. Moreover, we find a specific set of parameters for which the SPP does not move when it is half coated, and moves in the direction of the catalytic cap or away from it, when it is less than or more than half coated, respectively. We also introduce a method to quickly determine the velocity dependence of a SPP for small perturbations of the molecule-surface interaction potential for low P{\'e}clet numbers. Finally, we briefly consider the sensitivity of the self-propulsion velocity on the details of the local reactivity of the catalytic cap and we find that for reasonable parameters our results are robust with respect to sizable changes. Our findings have significant implications for the realization of SPP-based applications and interpretation of experimental results. 

The remainder of the document is structured as follows. In Section~\ref{sec:model} we introduce our multi-component model for bulk fluids. The boundary conditions required to achieve self-propulsion for arbitrary shapes are discussed in Section~\ref{sec:sdmot}. This general discussion is exemplified on the basis of a spherical Janus SPP suspended in hydrogen peroxide solution in Section~\ref{sec:sd_sphere}. In Section~\ref{sec:slip_model} the dilute limit of the multi-component self-propulsion model is analyzed and the slip model is recovered. This analysis is built upon in Section~\ref{sec:perturb}, where for low P{\'e}clet numbers, we introduce a method to determine the influence of small changes in the molecule-surface interaction potential on the self-propulsion velocity. We subsequently present our results in Section~\ref{sec:result}, which is partitioned into several subsections. We introduce our system parameters and numerical solution approach in Section~\ref{sub:paramet}. This is followed by a comparison between the full multi-component result and the dilute-limit slip model result for the motion of a colloidal SPP in Section~\ref{sub:compare}. We then analyze the nature of the flow field around the SPP and the dependence of its speed on the catalytic surface coverage in Section~\ref{sub:single}. This analysis is extended upon in Section~\ref{sub:hetero}, where we consider the effect of heterogeneities in the surface-molecule interaction on the SPP's speed. Our last result is presented in Section~\ref{sub:reaction}, wherein we briefly consider the influence of variations in the reactivity of the catalytic coating.  Finally, we summarize and discuss our results in Section~\ref{sec:disc} and present an outlook.

\section{\label{sec:model}The Multi-Component Model for Bulk Fluid}

In the following, we consider a three component mixture, consisting of hydrogen peroxide H$_{2}$O$_{2}$ (fuel), oxygen O$_{2}$ (reactant), and water H$_{2}$O (medium + reactant). However, in this paragraph we will start with a more general formulation of the multi-component model. 

\subsection{\label{sub:remark}Remarks on Multi-Component Modeling}

The choice of a multi-component description for the fluid is related to the composition of the `total fluid' typically used in experiments, which contains at least three species in the case of a hydrogen-peroxide solution. In addition, most current theoretical descriptions do not take properly into account the differences in the molecular masses of the species. Finally, at the high concentrations of fuel used in experiment, typically up to 10\% b.v. of H$_{2}$O$_{2}$,~\cite{Brown,Ebbens,Ebbens_12,Howse_07,Valadares_10,simmchen} make it unclear how the incompressibility constraint on the total fluid is preserved within the diffusion-advection formulation of the behavior of the individual species.

An accurate description of the physics in a multi-component system uses incompressibility as a constraint that alters the behavior of the individual components. The Maxwell-Stefan multi-component formalism is such a model, that describes a total fluid that satisfies the Navier-Stokes equations, comprised of separate species which satisfy complex diffusion-advection relations that ensure conservation of local fluid density.~\cite{Bird,Curtiss,Groot_Mazur} This model is, however, exceedingly complex and leaves many free parameters. For instance, the values of the species' cross-diffusion coefficients are typically not known and not measurable in experiments. In this section, we therefore present a reduced multi-component model that takes into account some of the complexities in this type of system, whilst leaving the number of free parameters to a minimum.

We assume a bulk liquid of infinite extent; boundary conditions will be imposed at a later stage. All components that comprise the fluid are fully miscible and no demixing takes place, for example, dissolved oxygen cannot cross the solvation threshold to form bubbles. The latter would require, e.g., the addition of a Cahn-Hilliard-type description of the mixing-demixing transition, which needlessly complicates the system for our purposes. 

\subsection{\label{sub:incomp}Incompressibility and the Fluid Density Profile}

In the following we will consider a multi-component fluid for which the mass and momentum transport is described using the Navier-Stokes equations. For any flow characterized by the Navier-Stokes formalism (both compressible and incompressible), the continuity equation of the fluid is given by 
\begin{equation}
\label{eq:fluid_cont} \frac{\partial}{\partial t} \rho(\mathbf{r}) + \mathbf{\nabla} \cdot \left( \rho(\mathbf{r}) \mathbf{u} (\mathbf{r}) \right) = 0,
\end{equation}
where $\rho(\mathbf{r})$ is the mass density, $\partial/\partial t$ denotes partial differentiation with respect to time, $\mathbf{\nabla}$ is the spatial differential operator, $\cdot$ denotes the dot product (such that $\nabla \cdot$ is the divergence), $\mathbf{u}(\mathbf{r})$ is the advective velocity, and the time dependence of the quantities is implicit. The incompressibility condition is given by 
\begin{equation}
\label{eq:incompressible} \mathbf{\nabla} \cdot \mathbf{u}(\mathbf{r}) = 0,
\end{equation}
which in combination with Eq.~\cref{eq:fluid_cont} implies that equivalently
\begin{equation}
\label{eq:equivalence}  \frac{\partial}{\partial t} \rho(\mathbf{r}) = - \mathbf{u}(\mathbf{r}) \cdot \mathbf{\nabla} \rho(\mathbf{r}) .
\end{equation}
This equivalence leads to a subtlety in the definition of incompressibility. Note that $(\partial/\partial t)\rho(\mathbf{r}) = 0$ and $\mathbf{\nabla} \rho(\mathbf{r}) = \mathbf{0} $ do not have to be satisfied simultaneously, only Eq.~\cref{eq:equivalence} has to hold for a flow field to be incompressible. That is, while the flow of a material can be incompressible, the material itself does not need to have a homogeneous density distribution.

In the following we will choose to impose a homogeneous density distribution of the material, i.e., $\rho \equiv \rho(\mathbf{r})$. This is a reasonable assumption since we can prepare our system with a homogeneous density distribution and our boundary conditions all conserve local density (reactions are mass conserving), as we will see. Therefore, in the initial state we find $\mathbf{\nabla} \rho(\mathbf{r}) = \mathbf{0}$; consequently incompressibility guarantees (through $(\partial/\partial t)\rho(\mathbf{r}) = 0$) that the density remains homogeneous. In the derivation below this condition may be relaxed, however, the final result will not be substantially affected.

\subsection{\label{sub:chem}Chemical Potential and Diffusive Flux}

We start our derivation from the expression for the chemical potential~\cite{DeHoff_06} of the species that comprise a non-ideal, multi-component mixture
\begin{equation}
\label{eq:chem_pot} \mu_{k}(\mathbf{r}) = \mu_{k}^{0} + k_{\mathrm{B}}T \log \gamma_{k}(\mathbf{r},\{x_{l}\}) + k_{\mathrm{B}}T \log x_{k}(\mathbf{r}) + \Phi_{k}(\mathbf{r}) ,
\end{equation}
where the subscript $k$ is used to identify the $k$-th species out of $N$ total species; $\mu_{k}^{0}$ is the reference chemical potential of that species; $k_{\mathrm{B}}$ is the Boltzmann constant and $T$ is the temperature; $\gamma_{k}(\mathbf{r},\{x_{l}\})$ is the activity coefficient of species $k$ with $\mathbf{r}$ the position coordinate; $\Phi_{k}(\mathbf{r})$ is an external potential acting on molecules of species $k$; and $x_{k}(\mathbf{r})$ is the molar fraction of species $k$, which is defined as
\begin{equation}
\label{eq:x_def} x_{k}(\mathbf{r}) \equiv \frac{n_{k}(\mathbf{r})}{\sum_{j=1}^{N} n_{j}(\mathbf{r})},
\end{equation}
with $n_{k}(\mathbf{r})$ the particle density of species $k$. The activity coefficient accounts for non-ideality and is therefore dependent on the concentrations of the other species as well, which is denoted here by the use of the set $\{x_{l}\}$. In the following we specify $\gamma_{k}(\mathbf{r},\{x_{l}\})$ to be constant in $\mathbf{r}$; this will necessitate further approximations to be made in order to recover incompressibility of the total fluid in the following. The first strong reduction is therefore made here, as the fluid is considered an ideal mixture; effectively we set $\gamma_{k}(\mathbf{r},\{x_{l}\}) = 1$.

In order to derive the continuity equations for the $k$-th species, we need to introduce several quantities. The mass density of the total fluid is given by $\rho$ and is the sum of the component's mass densities $\rho_{k}(\mathbf{r})$. We may now introduce the mass fractions of the species as
\begin{equation}
\label{eq:mfrac} \omega_{k}(\mathbf{r}) \equiv \frac{\rho_{k}(\mathbf{r})}{\rho}.
\end{equation}
Species $k$ has a molecular mass of $m_{k}$, which gives us the equivalence $\rho_{k}(\mathbf{r}) = m_{k} n_{k}(\mathbf{r})$ and allows us to write 
\begin{equation}
\label{eq:maver} m(\mathbf{r}) \equiv \left( \sum_{k=1}^{N} \frac{\omega_{k}(\mathbf{r})}{m_{k}} \right)^{-1},
\end{equation}
if we assume that there is a field $m(\mathbf{r})$ such that $\rho = m(\mathbf{r})n(\mathbf{r})$, with $n(\mathbf{r})$ the total number density. Here, $m(\mathbf{r})$ is a quantity that has the properties of a number-density averaged molecular mass. That is, an average molecular mass of the fluid, based on the composition of the fluid into species with different molecular masses. Note that if all molecular masses are equal, then $m(\mathbf{r})$ is position (and time) independent, since the sum over $\omega_{i}(\mathbf{r})$ is unity by definition. Because we are interested in an incompressible total fluid, it is convenient to recast the expression for the molar fractions in terms of the above quantities
\begin{equation}
\label{eq:x_rho} x_{k}(\mathbf{r}) = \frac{\omega_{k}(\mathbf{r}) m(\mathbf{r})}{m_{k}}.
\end{equation}
This results in the following expression for the gradient (denoted by $\mathbf{\nabla}$) of the molar fraction
\begin{equation}
\label{eq:Dx} \mathbf{\nabla} x_{k}(\mathbf{r}) = \frac{m(\mathbf{r})}{m_{k}} \mathbf{\nabla} \omega_{k}(\mathbf{r}) + \frac{\omega_{k}(\mathbf{r}) }{m_{k}} \mathbf{\nabla} m(\mathbf{r}).
\end{equation}

Using the above quantities and Eq.~\cref{eq:Dx}, it follows that the expression for the number density flux $\mathbf{J}^{\ast}_{k}(\mathbf{r})$ of species $k$, in the frame of reference that is co-moving with the total fluid, can be written as
\begin{eqnarray}
\label{eq:J_def} \mathbf{J}^{\ast}_{k}(\mathbf{r}) & = & \nu_{k} n_{k}(\mathbf{r}) \mathbf{f}_{k}(\mathbf{r}) \\
\label{eq:J_defplug} & = & -\nu_{k} n_{k}(\mathbf{r}) \mathbf{\nabla} \mu_{k}(\mathbf{r})  \\
\nonumber & = & -\frac{ D_{k} n_{k}(\mathbf{r})}{k_{\mathrm{B}}T} \mathbf{\nabla} \mu_{k}(\mathbf{r}) ; \\
\nonumber & = & - \frac{ D_{k} \rho }{m_{k}} \bigg( \mathbf{\nabla} \omega_{k}(\mathbf{r}) + \omega_{k}(\mathbf{r}) \frac{\mathbf{\nabla} m(\mathbf{r}) }{m(\mathbf{r})} \\
\label{eq:J_fin} & & \phantom{ - \frac{ D_{k} \rho}{m_{k}} \bigg( \mathbf{\nabla} \omega_{k}(\mathbf{r}) } + \omega_{k}(\mathbf{r}) \mathbf{\nabla} \tilde{\Phi}_{k}(\mathbf{r}) \bigg),
\end{eqnarray}
where $\nu_{k}$ is the mobility of the $k$-th species, which is related to that species' diffusion constant $D_{k}$ via the Einstein-Smoluchowski relation $\nu_{k} = D_{k}/k_{\mathrm{B}}T$; $\mathbf{f}_{k}(\mathbf{r})$ is the force per particle acting on the component; and $\tilde{\Phi}_{k}(\mathbf{r}) \equiv \Phi_{k}(\mathbf{r})/k_{\mathrm{B}}T$. Here, we used the first order expansion for small deviations of $\mu_{k}(\mathbf{r})$ from $\mu_{k}^{0}$, in order to write $\mathbf{f}_{k}(\mathbf{r}) = - \mathbf{\nabla}\mu_{k}(\mathbf{r})$ in Eq.~\cref{eq:J_defplug}.~\cite{Groot_Mazur} 

\subsection{\label{sub:mass_flux}Mass Flux and Continuity}

The expression in Eq.~\cref{eq:J_fin} allows us to write the kinematic mass flux (in the co-moving frame) as
\begin{eqnarray}
\nonumber \mathbf{j}^{\ast}_{k}(\mathbf{r}) & = & \frac{m_{k}\mathbf{J}^{\ast}_{k}}{\rho} \\
\nonumber & = & -D_{k} \left( \mathbf{\nabla} \omega_{k}(\mathbf{r}) + \omega_{k}(\mathbf{r}) \frac{\mathbf{\nabla}m(\mathbf{r})}{m(\mathbf{r})} + \omega_{k}(\mathbf{r}) \mathbf{\nabla} \tilde{\Phi}_{k}(\mathbf{r}) \right). \\
\label{eq:fluxsp} & &
\end{eqnarray}
Note that the kinematic mass flux has the physical dimension of velocity. Further note that when all species have the same molecular mass, the mass-correction (second) term drops out of the above equations, which allows us to recover the traditional Fickian diffusion. 

The kinematic mass flux in the co-moving frame $\mathbf{j}^{\ast}_{k}(\mathbf{r})$ and the kinematic mass flux in the laboratory frame $\mathbf{j}_{k}(\mathbf{r})$ are related according to the following transformation
\begin{equation}
\label{eq:jastdefined} \mathbf{j}_{k}(\mathbf{r}) = \mathbf{j}_{k}^{\ast}(\mathbf{r}) + \omega_{k}(\mathbf{r}) \mathbf{u}(\mathbf{r}),
\end{equation}
where $\mathbf{u}(\mathbf{r})$ is the velocity of the total fluid, which accounts for the advective contribution. The time-dependent continuity equations for a system, in which there are no bulk reactions, read
\begin{equation}
\label{eq:contlab} \frac{\partial}{\partial t}{\omega}_{k}(\mathbf{r}) + \mathbf{\nabla} \cdot \mathbf{j}_{k} (\mathbf{r}) = 0.
\end{equation}
Note that bulk reaction terms may be straightforwardly accounted for in our formalism by introducing source and sink terms into Eq.~\cref{eq:contlab}. The time-dependent continuity equation for the $k$-th species in the co-moving frame now follows from the relation in Eq.~\cref{eq:jastdefined}
\begin{equation}
\label{eq:contsp} \frac{\partial}{\partial t}{\omega}_{k}(\mathbf{r}) + \mathbf{\nabla} \cdot \mathbf{j}^{\ast}_{k} (\mathbf{r}) + \mathbf{u}(\mathbf{r})\cdot\mathbf{\nabla}\omega_{k}(\mathbf{r}) + \omega_{k}(\mathbf{r}) \mathbf{\nabla} \cdot \mathbf{u}(\mathbf{r}) = 0, 
\end{equation}
The above set of equations corresponds to the simplified diffusion-advection model put forward in Ref.~\cite{Bird}.

\subsection{\label{sub:momentum}Momentum Transport}

For momentum transport in the total fluid, we are interested in the low Reynolds number limit. That is, $Re \ll 1$, where $Re \equiv a u / \eta$, with $a$ the relevant length scale of the problem, $u$ the speed of the object or fluid, and $\eta$ the dynamic viscosity of the fluid. For colloidal SPPs typically used in experiments, $Re \ll 1$ is satisfied.~\cite{Brown,Ebbens,Ebbens_12,Howse_07,Valadares_10,simmchen} This implies that we can work with the linearized form of the Navier-Stokes equations for the total fluid
\begin{eqnarray}
\label{eq:NSred} \rho \frac{\partial}{\partial t} \mathbf{u}(\mathbf{r}) & = & - \mathbf{\nabla} p(\mathbf{r}) + \eta \mathbf{\Delta} \mathbf{u}(\mathbf{r}) + \mathbf{f}(\mathbf{r}) ; \\
\label{eq:fcont} \mathbf{\nabla} \cdot \mathbf{u}(\mathbf{r}) & = & 0, 
\end{eqnarray}
where $p(\mathbf{r})$ is the pressure, $\mathbf{\Delta}$ the tensorial Laplacian, and $\mathbf{f}(\mathbf{r})$ the force per volume acting on the fluid. Equation~\cref{eq:fcont} gives the incompressibility constraint for the total fluid. The force term $\mathbf{f}(\mathbf{r})$ derives from the particle fluxes of the individual species~\cite{Groot_Mazur}
\begin{equation}
\label{eq:f_origin} \mathbf{f}(\mathbf{r}) = \sum_{k=1}^{N} n_{k}(\mathbf{r}) \mathbf{f}_{k}(\mathbf{r}) = \sum_{k=1}^{N} \frac{\mathbf{J}^{\ast}_{k}(\mathbf{r})}{\nu_{k}} = \sum_{k=1}^{N} \frac{\rho \mathbf{j}^{\ast}_{k}(\mathbf{r})}{m_{k} \nu_{k}} .
\end{equation}
Using Eq.~\cref{eq:fluxsp} it now follows that
\begin{eqnarray}
\nonumber \mathbf{f}(\mathbf{r}) & = & - k_{\mathrm{B}} T \rho \left( \mathbf{\nabla} \frac{1}{m(\mathbf{r})} + \frac{1}{m(\mathbf{r})} \frac{\mathbf{\nabla} m(\mathbf{r})}{m(\mathbf{r})} \right) \\
\nonumber & & - k_{\mathrm{B}} T \sum_{k=1}^{N} \rho \frac{\omega_{k}(\mathbf{r})}{m_{k}}  \mathbf{\nabla} \tilde{\Phi}_{k}(\mathbf{r}) ; \\
\label{eq:f_result} & = & - k_{\mathrm{B}} T \sum_{k=1}^{N} \rho \frac{\omega_{k}(\mathbf{r})}{m_{k}} \mathbf{\nabla} \tilde{\Phi}_{k}(\mathbf{r}) .
\end{eqnarray}
The first two terms exactly cancel each other for an incompressible fluid. This is a nice feature of the multi-component model that takes into account the molecular mass of the species and utilizes incompressibility, which is missing in the traditional dilute-limit approximation. 

\subsection{\label{sub:incom_flux}Incompressibility and Flux}

The continuity equation of the total fluid, see Eq.~\cref{eq:fcont}, should also follow by summing the continuity equations of the individual species, see Eq.~\cref{eq:contsp}, in order for the model to be self-consistent in the incompressibility assumption. From the properties of $\omega_{k}(\mathbf{r})$ it follows that a sufficient condition to obtain incompressibility of the total fluid from the individual continuity equations, see Eq.~\cref{eq:contsp}, is given by
\begin{equation}
\label{eq:jcond_fin} \sum_{k=1}^{N} \mathbf{j}^{\ast}_{k}(\mathbf{r}) = \mathbf{0} .
\end{equation}
This is also physically reasonable as otherwise there would be mass flow with respect to the frame co-moving with the fluid (local center of mass). 

To ensure that Eq.~\cref{eq:jcond_fin} is satisfied, constraints on the species' fluxes must be imposed, as the current expressions do not necessarily result in incompressibility of the total fluid. In particular, there is insufficient coupling between the individual mass fractions $\omega_{k}(\mathbf{r})$ to ensure that this constraint is satisfied. This is typically the point where cross-diffusion terms are introduced in order to recover incompressibility.~\cite{Groot_Mazur,Bird,Curtiss} However, the introduction of cross-diffusion terms leads to complicated relations between the components and diffusion coefficients that cannot be (easily) extracted from experimental data. 

Here, we propose a different route to salvage the `simple' Fickian diffusion model. We first consider the force-free part of the flux. We may write this flux as 
\begin{equation}
\label{eq:fluxsp0} \mathbf{j}^{\ast,0}_{k}(\mathbf{r}) =  -D_{k} \left( \mathbf{\nabla} \omega_{k}(\mathbf{r}) + \omega_{k}(\mathbf{r}) \frac{\mathbf{\nabla}m(\mathbf{r})}{m(\mathbf{r})}  \right) .
\end{equation}
If we impose
\begin{equation}
\label{eq:fluxinc} \mathbf{j}^{\ast,0}_{1}(\mathbf{r}) \equiv -\sum_{k=2}^{N} \mathbf{j}^{\ast,0}_{k}(\mathbf{r}) ,
\end{equation} 
then we obtain 
\begin{equation}
\label{eq:jcond_0} \sum_{k=1}^{N} \mathbf{j}^{\ast,0}_{k}(\mathbf{r}) = \mathbf{0} .
\end{equation}
Here, we assumed that one of the components has a mass fraction that is substantially greater than the other components, namely the solvent. This component, say it is labeled $k = 1$, is singled out and employed to cancel momentum, flux, and density related inconsistencies, when there are no forces acting. We must still take care of contribution coming from the potentials. 

To ensure zero net flux in the co-moving frame, some correction force $\mathbf{f}_{\mathrm{corr}}(\mathbf{r})$ must be added to the gradients of the potentials. We make the following ansatz for the total force $\mathbf{f}_{\mathrm{tot},k}(\mathbf{r})$ acting on particles of the $k$-th species in the co-moving frame
\begin{equation}
\label{eq:ansatz} \mathbf{f}_{\mathrm{tot},k}(\mathbf{r}) = - \mathbf{\nabla} \tilde{\Phi}_{k}(\mathbf{r}) - \mathbf{f}_{\mathrm{corr}}(\mathbf{r}).
\end{equation}
By plugging in $\mathbf{f}_{\mathrm{tot},k}(\mathbf{r})$ into Eq.~\cref{eq:fluxsp} and utilizing Eq.~\cref{eq:fluxinc}, the following relation can be derived for the correction force such that Eq.~\cref{eq:jcond_fin} is satisfied and that incompressibility of the total fluid is recovered
\begin{equation}
\label{eq:fict} \mathbf{f}_{\mathrm{corr}}(\mathbf{r}) = - \frac{\sum_{k=1}^{N} D_{k} \omega_{k}(\mathbf{r}) \mathbf{\nabla} \tilde{\Phi}_{k}(\mathbf{r}) }{\sum_{k=1}^{N} D_{k} \omega_{k}(\mathbf{r})},
\end{equation}
We may thus write for the co-moving kinematic mass flux 
\begin{eqnarray}
\nonumber \mathbf{j}^{\ast}_{k>1}(\mathbf{r}) & = & -D_{k} \bigg( \mathbf{\nabla} \omega_{k}(\mathbf{r}) + \omega_{k}(\mathbf{r}) \frac{\mathbf{\nabla}m(\mathbf{r})}{m(\mathbf{r})} \\
\label{eq:jast} & & \phantom{-D_{k} \bigg(} - \omega_{k}(\mathbf{r}) \mathbf{f}_{\mathrm{tot},k}(\mathbf{r}) \bigg) ; \\
\label{eq:def0} \mathbf{j}^{\ast}_{1}(\mathbf{r}) & = & - \sum_{k=2}^{N} \mathbf{j}^{\ast}_{k}(\mathbf{r}) ; \\
\nonumber \mathbf{f}_{\mathrm{tot},k}(\mathbf{r}) & = & - \left( \mathbf{\nabla} \tilde{\Phi}_{k}(\mathbf{r}) - \frac{\sum_{k=1}^{N} D_{k} \omega_{k}(\mathbf{r}) \mathbf{\nabla} \tilde{\Phi}_{k}(\mathbf{r}) }{\sum_{k=1}^{N} D_{k} \omega_{k}(\mathbf{r})} \right) . \\
\label{eq:ftotk} & &
\end{eqnarray}
Note that Eq.~\cref{eq:def0} now follows from the previously imposed condition of Eq.~\cref{eq:fluxinc} and the properties of $\mathbf{f}_{\mathrm{tot},k}(\mathbf{r})$. 

\subsection{\label{sub:props_reds}Further Properties and Reductions}

The above flux model has the desirable property that throughout space, only the continuity equations for $N-1$ of the species have to be solved simultaneously, as is expected for an incompressible system. Moreover, the potential acting on the solvent is not ignored, as would be the case if the condition of Eq.~\cref{eq:def0} had been imposed without modification of the force acting on the fluid.  Note that this modification is only appropriate for the flux in the co-moving frame. The force acting on the fluid, see Eq.~\cref{eq:f_result}, should not be modified. This becomes clear by considering the case wherein all $D_{k}$ and $m_{k}$ are the same, for which imposing the correction would lead to a zero net force in the momentum transport equation.

Since we have shown that the continuity equation for the total fluid can be derived from the individual continuity equations of the species for the above flux expression, we can now split these two parts, writing
\begin{equation}
\label{eq:cont_fluid} \mathbf{\nabla} \cdot \mathbf{u}(\mathbf{r}) = 0
\end{equation}
for the fluid and 
\begin{equation}
\label{eq:cont_flux} \frac{\partial}{\partial t}{\omega}_{k}(\mathbf{r}) + \mathbf{\nabla} \cdot \mathbf{j}^{\ast}_{k} (\mathbf{r}) + \mathbf{u}(\mathbf{r})\cdot\mathbf{\nabla}\omega_{k}(\mathbf{r}) = 0, 
\end{equation}
for the diffusive species, respectively.

\subsection{\label{sub:over}Overview of the Model}

Summarizing, the multi-component model for the bulk fluid satisfies the following equations under the assumption of a homogeneous incompressible medium
\begin{eqnarray}
\nonumber \rho \frac{\partial}{\partial t} \mathbf{u}(\mathbf{r}) & = & - \mathbf{\nabla} p(\mathbf{r}) + \eta \mathbf{\Delta} \mathbf{u}(\mathbf{r}) \\
\label{eq:NS_sum} & & - k_{\mathrm{B}} T \rho \sum_{k=1}^{N} \frac{\omega_{k}(\mathbf{r})}{m_{k}} \mathbf{\nabla} \tilde{\Phi}_{k}(\mathbf{r}) ; \\
\label{eq:NS_cont} \mathbf{\nabla} \cdot \mathbf{u}(\mathbf{r}) & = & 0 ; \\
\nonumber \frac{\partial}{\partial t}{\omega}_{k}(\mathbf{r}) & = & - \mathbf{\nabla} \cdot \mathbf{j}^{\ast}_{k} (\mathbf{r}) - \mathbf{u}(\mathbf{r})\cdot\mathbf{\nabla}\omega_{k}(\mathbf{r}) ; \\
\label{eq:cont_sum} & & \\
\nonumber \mathbf{j}^{\ast}_{k>1}(\mathbf{r}) & = & -D_{k} \bigg( \mathbf{\nabla} \omega_{k}(\mathbf{r}) + \omega_{k}(\mathbf{r}) \frac{\mathbf{\nabla}m(\mathbf{r})}{m(\mathbf{r})} \\
\label{eq:jast_sum} & & \phantom{-D_{k} \bigg(} - \omega_{k}(\mathbf{r}) \mathbf{f}_{\mathrm{tot},k}(\mathbf{r}) \bigg) ; \\
\label{eq:def_sum} \mathbf{j}^{\ast}_{1}(\mathbf{r}) & = & - \sum_{k=2}^{N} \mathbf{j}^{\ast}_{k}(\mathbf{r}) ; \\
\label{eq:ftotk_sum} \mathbf{f}_{\mathrm{tot},k}(\mathbf{r}) & = & - \left( \mathbf{\nabla} \tilde{\Phi}_{k}(\mathbf{r}) - \frac{\sum_{k=1}^{N} D_{k} \omega_{k}(\mathbf{r}) \mathbf{\nabla} \tilde{\Phi}_{k}(\mathbf{r}) }{\sum_{k=1}^{N} D_{k} \omega_{k}(\mathbf{r})} \right) ; \quad \quad \\
\label{eq:ma_sum} m(\mathbf{r}) & = & \left( \sum_{k=1}^{N} \frac{\omega_{k}(\mathbf{r})}{m_{k}} \right)^{-1}.
\end{eqnarray}
This set of equations is fully consistent with the incompressibility assumption for the total fluid and dependent on a minimal number of fields: $\mathbf{u}(\mathbf{r})$, $p(\mathbf{r})$, $\omega_{k}(\mathbf{r})$, and $\tilde{\Phi}_{k}(\mathbf{r})$; and parameters: $T$, $\eta$, $\rho$, $D_{k}$, and $m_{k}$.

Note that this model presents a significant departure from the typical dilute-limit approximation, since the difference in mass of the molecular species is taken into account and incompressibility is incorporated in a self-consistent manner. However, our modifications and underlying assumptions require the solute species to remain dilute in the solvent. It is therefore not of the same class as the more realistic, albeit far more complex Maxwell-Stefan formulation for higher solute concentrations. 

\section{\label{sec:sdmot}Self-Diffusiophoretic Motion of Objects of Arbitrary Shape}

Thus far, we have only concentrated on the description of the `bulk' fluid, not the mechanism by which self-propulsion is achieved. To obtain self-diffusiophoretic motion of a particle in our multi-component description we require only a set of boundary conditions. We are interested in the stationary state of the system and we therefore solve the time-independent variants of Eqs.~(\ref{eq:NS_sum}-\ref{eq:ma_sum}). By solving these equations simultaneously, the self-propulsion velocity and induced fluid flow profile in the stationary state may be obtained.

At the surface of the particle we require a no-slip boundary condition. That is, 
\begin{equation}
\label{eq:bc_u} \mathbf{u}(\mathbf{s}) = 0,
\end{equation}
where $\mathbf{s}$ is a point on the surface. We also require a flux boundary condition on (part of) the surface to cause the system to go out of equilibrium. For a multi-component system there can be a large number of (non-linear) reactions taking place simultaneously between the surface and the various components. The only requirement on the reactions is that there is no net mass flux into the surface
\begin{equation}
\label{eq:bc_j} \sum_{k=1}^{N} \mathbf{j}_{k}\left( \mathbf{s} \right) \cdot \hat{\mathbf{n}}(\mathbf{s}) = 0, 
\end{equation} 
where $\hat{\mathbf{n}}(\mathbf{s})$ is the normal vector to the surface at the point $\mathbf{s}$. If there were a net mass flux into the surface, the no-slip boundary condition of Eq.~\cref{eq:bc_u} would be violated, due to the relation between $\mathbf{j}_{k}(\mathbf{r})$ and $\mathbf{u}(\mathbf{r})$. In physical terms, the condition of Eq.~\cref{eq:bc_j} imposes that the reactions are mass conserving. The choice for the reaction scheme is otherwise completely free. 

There are a few caveats to the formulation of the system we presented above.
\begin{itemize}
\item There is \emph{no} fluid motion in the stationary state when all $\tilde{\Phi}_{k}(\mathbf{r}) = 0$. That is, in the absence of an interaction between the surface of the SPP and the solvent/solutes, there is no back-coupling to the total fluid of the heterogeneous solute distribution caused by the surface reaction. This leads to zero fluid flow, in accordance with the results of Ref.~\cite{Brady_11}. 
\item Solving the time-independent variants of Eqs.~(\ref{eq:NS_sum}-\ref{eq:ma_sum}) requires the existence of a stationary state, for which there is an inertial transformation between the frame in which the particle moves and the frame in which the particle is stationary and the fluid moves. 
\item The existence of a inertial transformation to a co-moving frame implies that the particle performs rectilinear motion that is unaccelerated. This imposes restrictions on the shape of the particle and surface coating heterogeneity. Namely, the particle must be such that it does not experience a torque. A sufficient condition is that the particle is rotationally symmetric in the shape and surface coating. Moreover, there are restrictions on the reaction mechanisms that are permitted, e.g., the condition of stationarity forbids oscillating reactions.
\item Finally, some care needs to be taken in setting up the boundary conditions far away from the colloid, to ensure that a stationary state with non-zero self-propulsion velocity may be achieved, i.e., fuel must not be depleted.
\end{itemize}

\section{\label{sec:sd_sphere}A Self-Diffusiophoresing Sphere in Hydrogen Peroxide}

In this section, we concentrate exclusively on a single spherical SPP of radius $a$, that is suspended in a three-component mixture of water, hydrogen peroxide, and oxygen ($N = 3$), to better illustrate our auto-phoretic model. This colloid is partially coated with platinum to allow for a catalytic decomposition reaction of the hydrogen peroxide to take place; the uncoated part of the particle is assumed to be nonreactive. The sphere is centered in and fixed to the origin of our coordinate system. We use spherical coordinates with $r$ the distance measured to the origin, $\theta$ the azimuthal angle, and $\phi$ the polar angle. The system is rotationally symmetric around the $z$-axis and $\phi$ is measured with respect to this axis, whereas $\theta$ describes rotations around it. Let the normal vector to the sphere's surface (radial unit vector) be denoted by $\hat{\mathbf{n}}$ and the polar-angle tangent vector by $\hat{\mathbf{t}}$; the Cartesian unit vectors are given by $\hat{\mathbf{x}}$, $\hat{\mathbf{y}}$, and $\hat{\mathbf{z}}$.

Using these notations the boundary conditions at the surface of the sphere are as follows. We include a flux boundary condition to describe a simple linear reaction 2~H$_{2}$O$_{2}$ $\rightarrow$ 2~H$_{2}$O $+$ O$_{2}$ for hydrogen peroxide decomposition at the platinum surface. The reaction kinetics satisfy
\begin{eqnarray}
\label{eq:h2o2} \frac{\partial}{\partial t} \left[ \mathrm{H}_{2}\mathrm{O}_{2} \right] & \stackrel{\mathrm{Pt}}{=} & - k \left[ \mathrm{H}_{2}\mathrm{O}_{2} \right] ; \\
\label{eq:h2o} \frac{\partial}{\partial t} \left[ \mathrm{H}_{2}\mathrm{O} \right] & \stackrel{\mathrm{Pt}}{=} & -\frac{\partial}{\partial t} \left[ \mathrm{H}_{2}\mathrm{O}_{2} \right] ; \\
\label{eq:o2} \frac{\partial}{\partial t} \left[ \mathrm{O}_{2} \right] & \stackrel{\mathrm{Pt}}{=} & -\frac{1}{2} \frac{\partial}{\partial t} \left[ \mathrm{H}_{2}\mathrm{O}_{2} \right] ,
\end{eqnarray}
where $k$ is the reaction rate. The set of equations, Eqs.~(\ref{eq:h2o2}-\ref{eq:o2}), is used to treat bulk catalytic decomposition and needs to be converted to a surface model, considering the specific geometrical properties of the thin Pt-coating of the SPP, which we will do shortly. Other choices for the reaction mechanism by which hydrogen peroxide is decomposed at a platinum surface are possible, as set out in Ref.~\cite{Sabass_12b}. A multiple-step reaction mechanism through a variety of intermediate Pt-complexes is a more probable route for the decomposition.~\cite{Hall1,Hall2} However, there is some indication that the rate-limiting step in the reaction process is linear in the hydrogen-peroxide concentration.~\cite{Sabass_12b} Taking the above considerations into account, we may write for the flux boundary conditions
\begin{equation}
\label{eq:reacbc} \left. \mathbf{j}_{i}(\mathbf{r}) \cdot \hat{\mathbf{n}} \right\vert_{r=a} = \sum_{k=1}^{3} R(\phi) a_{ik} \left. \omega_{k}(\mathbf{r}) \right\vert_{r=a},
\end{equation}
where the $a_{ik}$ are stoichiometric/mass coefficients and $R(\phi)$ is the surface reaction rate. The relation to the bulk reaction rate $k$ is given by $R(\phi) = k(\phi)/S_{c}$, where $k(\phi)$ is the bulk reaction rate and $S_{c}$ is the catalytic surface area for which this rate is achieved. The $\phi$-dependence of the reaction rate allows us to treat one part of the SPP's surface as inert and one as catalytic. The stoichiometric/mass coefficients are conveniently summarized using the following matrix notation
\begin{equation}
\label{eq:reacco} A = [a_{ij}] = \left( \begin{array}{ccc} \displaystyle \frac{m_{\mathrm{H}_{2}\mathrm{O}}}{m_{\mathrm{H}_{2}\mathrm{O}_{2}}} & 0 & 0 \\[12pt] -1 & 0 & 0 \\[3pt] \displaystyle \frac{m_{\mathrm{O}_{2}}}{2 m_{\mathrm{H}_{2}\mathrm{O}_{2}}} & 0 & 0 \end{array} \right) ,
\end{equation}
where we used the assignment $i = 1$ (H$_{2}$O), $i = 2$ (H$_{2}$O$_{2}$), and $i = 3$ (O$_{2}$). Note that the two zero columns imply that only a single decomposition reaction takes place.

A no-slip boundary condition is imposed at the surface of the particle
\begin{equation}
\label{eq:bcvelo_surf} \left. \mathbf{u}(\mathbf{r}) \right\vert_{r=a} = \mathbf{0},
\end{equation}
Finally, we set the following boundary conditions at `infinity', the edge of our simulation/computation domain. The mass fractions of the various species are set to their `reservoir' values
\begin{equation}
\label{eq:reserv} \left. \omega_{i}(\mathbf{r}) \right\vert_{r \uparrow \infty} = \omega_{i,\mathrm{res}},
\end{equation}
where $\omega_{i,\mathrm{res}}$ is the reservoir mass fraction. This ensures the formation of a stationary state, i.e., there is no depletion of species as there is a continuous influx of fuel and outflow of products at the boundary. For the total fluid, the pressure boundary condition is 
\begin{equation}
\label{eq:presbc} \left. p(\mathbf{r}) \right\vert_{r \uparrow \infty} = 0,
\end{equation}
and the velocity boundary condition reads
\begin{equation}
\label{eq:bcvelo} \left. \mathbf{u}(\mathbf{r}) \right\vert_{r \uparrow \infty} = \mathbf{v},
\end{equation}
where $\mathbf{v}$ is the self-propulsion velocity of the colloid, for which we solve. Note that Eq.~\cref{eq:bcvelo} is a direct consequence of being in the frame co-moving with the SPP.

\section{\label{sec:slip_model}The Dilute Limit Slip Model for Self-Diffusiophoresis}

Before we move on to the results let us return to an $N$ component model. Solving Eqs.~(\ref{eq:NS_sum}-\ref{eq:ma_sum}) simultaneously is a non-trivial task, even for the simple system of a partially catalytic sphere. In literature the slip model is often used to allow for analytic treatment of the self-diffusiophoretic properties of (spherical) SPPs.~\cite{Anderson,Golestanian_07,Brady_08,Popescu_10,Brady_11,Ebbens_12,Mood_13,Brady_13,Brady_14,Lauga_14} Here, we show the additional assumptions that are required to arrive at an analytically tractable slip model from our set of multi-component equations. 

The slip model is based on an expansion of Eqs.~(\ref{eq:NS_sum}-\ref{eq:ma_sum}) in the small layer around the colloid where the molecule-surface interaction potential is non-negligible. Let $\delta$ be the `range' of the longest-ranged surface-molecule interaction potential. That is, for $\vert \mathbf{r} - \mathbf{s} \vert > \delta$ we have $\Phi_{k}(\mathbf{r}) \ll k_{\mathrm{B}}T$, where $\mathbf{s}$ is the closest point on the surface to $\mathbf{r}$. Let $\kappa^{-1}(\mathbf{s})$ be the local curvature, then we can write $\lambda(\mathbf{s}) = \delta \kappa(\mathbf{s})$. Note that $\kappa^{-1}(\mathbf{s}) = a$ for a sphere. We further introduce the concept of the local Damk{\"o}hler number, which gives the ratio of the reaction rate and the diffusive mass transfer rate
\begin{equation}
\label{eq:damkohler} Da_{k}(\mathbf{s}) = \frac{R(\mathbf{s})}{D_{k}\kappa(\mathbf{s})},
\end{equation}
where $R(\mathbf{s})$ is the reaction rate. In order to apply the slip-layer approximation, an expansion of Eqs.~(\ref{eq:NS_sum}-\ref{eq:ma_sum}) in terms of $\lambda(\mathbf{s})$ and $Da(\mathbf{s})$, must be accurate to first order. This implies that $\lambda(\mathbf{s}) \ll 1$ and $Da_{k}(\mathbf{s}) \ll 1$, for all points on the surface and for all $k$. Sharifi-Mood~\emph{et al.}~\cite{Mood_13} investigated the validity of the first-order approximation and concluded that its range of applicability depends strongly on the nature of the interaction potential and the size of the colloid, however, $\lambda(\mathbf{s}) < 10^{-3}$ would be in the right regime.

When the first-order approximation is valid, we obtain local, instantaneous equilibrium in the small layer around the particle in the direction perpendicular to the surface. For all $k$ the following holds
\begin{equation}
\label{eq:slass} \mathbf{j}^{\ast}_{k} (\mathbf{r}) \cdot \hat{\mathbf{n}}(\mathbf{s}) = 0 ,
\end{equation}
where $\mathbf{s}$ is the closest point to the surface of the particle.~\cite{Brady_11} For convenience, we introduce the notation $z = \vert \mathbf{r} - \mathbf{s} \vert$ for the distance perpendicular to the surface. Then Eq.~\cref{eq:slass} can be written as
\begin{equation}
\label{eq:eqj} j^{\ast}_{k} (z) \equiv \mathbf{j}^{\ast}_{k} (\mathbf{s} + z \hat{\mathbf{n}}(\mathbf{s})) \cdot \hat{\mathbf{n}}(\mathbf{s}) = 0 .
\end{equation}
Using the assumption of Eq.~\cref{eq:eqj} the $\omega_{k}(z) \equiv \omega_{k} (\mathbf{s} + z \hat{\mathbf{n}}(\mathbf{s}))$ are solved within the layer to obtain the species profiles perpendicular to the surface. However, since there are coupling terms in our expressions for $j^{\ast}_{k} (z)$, this is a non-trivial problem.

If we assume that $\mathbf{\nabla}m(\mathbf{r}) \approx 0$, the equations for the kinematic mass fluxes decouple and the above condition reduces to
\begin{equation}
\label{eq:jastred} j^{\ast}_{k}(z) = -D_{k} \left( \frac{\partial}{\partial z} \omega_{k}(z) - \omega_{k}(z) F_{\mathrm{tot},k}(z) \right) = 0,
\end{equation}
which can be solved analytically for $\omega_{k}(z)$, with $F_{\mathrm{tot},k}(z) \equiv \mathbf{F}_{\mathrm{tot},k} (\mathbf{s} + z \hat{\mathbf{n}}(\mathbf{s})) \cdot \hat{\mathbf{n}}(\mathbf{s})$. Note that this implies that the molecular masses are roughly equal. The solution can be written as
\begin{equation}
\label{eq:omegz} \omega_{k}(z) = \omega_{k}(\delta) \exp \left( \int_{\delta}^{z} F_{\mathrm{tot},k}(z') \ud z'  \right) ,
\end{equation}
where it is assumed that $F_{\mathrm{tot},k}(r>\delta) = 0$. We still require
\begin{equation}
\label{eq:omegaz1} \omega_{1}(z) = 1 - \sum_{k=2}^{N} \omega(z).
\end{equation}
This solution can be inserted into the momentum balance of the Stokes' equation normal to the surface, which (up to first order, see Ref.~\cite{Brady_11}) reads
\begin{equation}
\label{eq:stokesz} \frac{\partial}{\partial z} p(z) = - k_{\mathrm{B}} T \rho \sum_{k=1}^{N} \frac{\omega_{k}(z)}{m_{k}} \frac{\partial}{\partial z} \tilde{\Phi}_{k}(z),
\end{equation} 
where $p(z) \equiv p(\mathbf{s} + z \hat{\mathbf{n}}(\mathbf{s}))$. We may solve this equation for the pressure decay perpendicular to the surface. Note that we can pull the $\tilde{m} \equiv m_{k}$ out of the sum, since we had already assumed the difference between the molecular masses to be small in establishing Eq.~\cref{eq:jastred}. However, by plugging the expression of Eq.~\cref{eq:omegz} into Eq.~\cref{eq:stokesz}, the system is still not analytically tractable, due to the complex form of $F_{\mathrm{tot},k}(z)$. 

We therefore make the additional assumptions that the diffusion coefficients are similar $\tilde{D} \equiv D_{k}$ and that $\omega_{1}(z) \gg \omega_{k>1}(z)$, which implies that $\omega_{1}(z) \approx 1$. This is exactly the dilute-limit approximation. Then we may write
\begin{eqnarray}
\label{eq:Ftotkdil} F_{\mathrm{tot},k}(z) & = & - \left( \frac{\partial}{\partial z} \tilde{\Phi}_{k}(z) - \frac{\partial}{\partial z} \tilde{\Phi}_{1}(z) \right) ; \\
\label{eq:phstdil} \tilde{\Phi}^{\ast}_{k}(z) & \equiv & \tilde{\Phi}_{k}(z) - \tilde{\Phi}_{1}(z) ;\\
\label{eq:omega_dil_new} \Rightarrow  \omega_{k}(z) & = & \omega_{k}(\delta) \exp \left( - \tilde{\Phi}^{\ast}_{k}(z) \right) ; \\
\nonumber \frac{\partial}{\partial z} p(z) & = & - k_{\mathrm{B}} T \frac{\rho}{\tilde{m}} \left( \frac{\partial}{\partial z} \tilde{\Phi}_{1}(z) + \sum_{k=2}^{N}  \omega_{k}(z) \frac{\partial}{\partial z} \tilde{\Phi}^{\ast}_{k}(z) \right) ; \\
\label{eq:stokeszdil} & & \\
\nonumber p(z) & = & - k_{\mathrm{B}} T \frac{\rho}{\tilde{m}} \bigg( \tilde{\Phi}_{1}(z) \\
\label{eq:psol} &   & \phantom{- k_{\mathrm{B}} T \frac{\rho}{\tilde{m}} \bigg(} + \sum_{k=2}^{N}  \omega_{k}(\delta) \left( e^{-\tilde{\Phi}^{\ast}_{k}(z)} - 1 \right) \bigg) ,
\end{eqnarray}
where we introduced the `effective' potential $\tilde{\Phi}^{\ast}_{k}(z)$, which is the potential of species $k$ with respect to that of the solvent. In addition, we assumed that $\tilde{\Phi}^{\ast}_{k}(\delta) = 0$ in writing Eq.~\cref{eq:omega_dil_new}. Equation~\cref{eq:psol} follows from Eq.~\cref{eq:phstdil} by making use of the boundary conditions at infinity. Our expression differs from the ones given by Sharifi-Mood~\emph{et al.}~\cite{Mood_13}, as the surface-solvent potential is not weighted by the ratio of the molecular masses. We consider the expression derived above to be the correct form for the effective potential in the dilute limit.

The solution for the pressure can be plugged into the Stokes equation for motion parallel to the surface.~\cite{Brady_11} Say that $y$ is the coordinate parallel to the surface and $v(y)$ is the velocity parallel to the surface, then
\begin{equation}
\label{eq:stokesy} \eta \frac{\partial^{2}}{\partial z^{2}} v(y) - \frac{\partial}{\partial y} p(y) = 0 .
\end{equation}
Now we make use of the fact that only $\omega_{k}(\delta)$ has a $y$ dependence (to first order) in Eq.~\cref{eq:psol}. The velocity of the fluid at a distance $\delta$ away from the surface is therefore given by
\begin{equation}
\label{eq:vy_del} v_{\delta}(y) = - \frac{k_{\mathrm{B}} T \rho}{\eta \tilde{m}} \sum_{k=2}^{N} \left( \frac{\partial}{\partial y} \omega_{k}(\delta) \right) \int_{0}^{\delta} s \left( e^{-\tilde{\Phi}^{\ast}_{k}(s)} - 1 \right) \ud s ,
\end{equation}
where terms of order $\mathcal{O}(\delta^{2})$ have been ignored in evaluating the partial integration required to arrive at the above form. Note that the above integral can be extended to infinity, since $\tilde{\Phi}^{\ast}_{k}(z) = 0$ for $z > \delta$. 

Combining the above results and translating them back to the three-dimensional (3D) result, we may now write for the boundary conditions at the edge of the slip layer
\begin{eqnarray}
\label{eq:unorm}\mathbf{u}\left(\mathbf{s} + \delta \hat{\mathbf{n}}(\mathbf{s}) \right)\cdot \hat{\mathbf{n}}(\mathbf{s}) & = & 0; \\
\nonumber \mathbf{u}\left(\mathbf{s} + \delta \hat{\mathbf{n}}(\mathbf{s}) \right)\cdot \hat{\mathbf{t}}(\mathbf{s}) & = &  - \sum_{k=2}^{N} \xi_{k}(\mathbf{s})  \\
\label{eq:slip} & & \phantom{- \sum_{k=2}^{N}} \times \hat{\mathbf{t}}(\mathbf{s}) \cdot \mathbf{\nabla} \omega_{k} \left(\mathbf{s} + \delta \hat{\mathbf{n}}(\mathbf{s}) \right) , \quad \quad 
\end{eqnarray}
where the interaction of the fluid molecules and the surface is taken into account by the coupling parameter
\begin{equation}
\label{eq:couple} \xi_{k}(\mathbf{s}) = \frac{ \rho k_{\mathrm{B}} T }{ \eta \tilde{m} } \int_{0}^{\infty} t \left( \exp \left( -\tilde{\Phi}^{\ast}_{k}(\mathbf{s} + t \hat{\mathbf{n}}(\mathbf{s}))\right) - 1 \right) \ud t .
\end{equation}
In Eq.~\cref{eq:slip}, $\hat{\mathbf{t}}(\mathbf{s})$ represents the tangent vectors to the surface at $\mathbf{s}$. N.B. There are two orthogonal tangent vectors at any two-dimensional (2D) surface in a 3D space. It is understood here that $\hat{\mathbf{t}}(\mathbf{s})$ is shorthand notation for both of these vectors. This makes Eq.~\cref{eq:slip} a set of two conditions, rather than a single condition. Further note that the above conditions are dependent on the position on the surface $\mathbf{s}$ and the length of the slip layer $\delta$. When $\delta \ll \kappa^{-1}(\mathbf{s})$, we may effectively take the limit $\delta \downarrow 0$ of Eqs.~(\ref{eq:unorm},\ref{eq:slip}) and assume that the slip velocity is imposed at the surface of the colloid.

\begin{figure}[!htb]
\begin{center}
\includegraphics[scale=1.0]{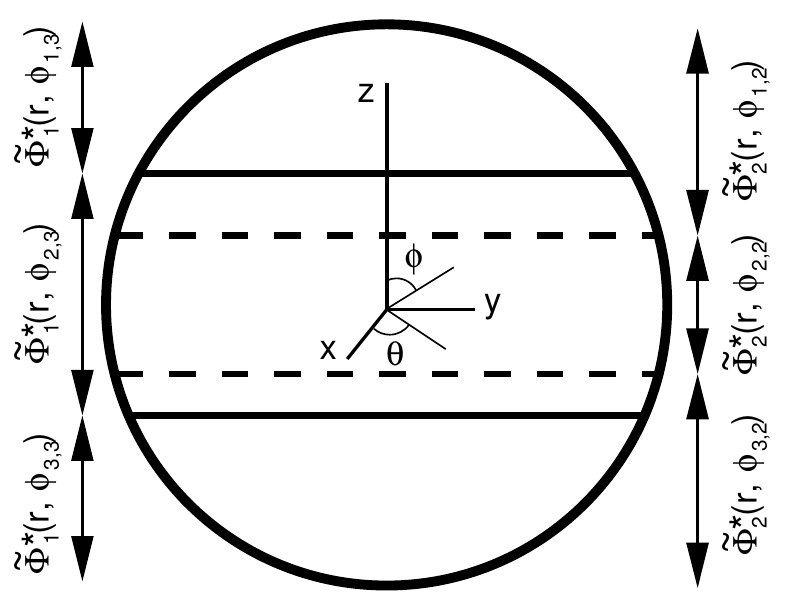}
\end{center}
\caption{\label{fig:bands}Sketch of a $z$-axisymmetric spherical particle, for which the surface is partitioned into stripes. On these stripes the effective molecule-surface interaction potential $\tilde{\Phi}^{\ast}_{k}(s,\phi)$ experienced by species $k$ is constant in $\phi$, the polar angle, but may have a distance dependence $r$. The azimuthal angle, in which the figure is symmetric, and the $x$- and $y$-axis are shown for completion. The stripes are identified by the label $\phi_{i,k}$, where $i$ refers to the index of the stripe and the subscript $k$ is used since the striped pattern need not coincide for different species. In this example two solute species are assumed; the stripes for species $k = 2$ are delimited using dashed lines and for species $k = 3$ using solid lines.}
\end{figure}

Finally, let us return to the spherical particles that we consider in this manuscript. N.B. Our spheres are assumed rotationally symmetric in the $z$-axis. The magnitude of the velocity of a spherical SPP that follows from the slip-model can be written in terms of the following surface integral
\begin{equation}
\label{eq:velocity} v = -\frac{1}{4\pi a^{2}} \sum_{k=2}^{N} \iint_{S} \xi_{k}(\phi) \left( \hat{\mathbf{t}}(\phi) \cdot \mathbf{\nabla} \omega_{k} (\phi) \right) \left( \hat{\mathbf{t}}(\phi) \cdot \hat{\mathbf{z}} \right) \ud \mathbf{s},
\end{equation}
where $S$ denotes the sphere's surface and $\ud \mathbf{s}$ is the measure for the surface integration.~\cite{Brady_11} The parameters $\hat{\mathbf{t}}(\phi)$, $\omega_{k} (\phi)$, and $\xi_{k}(\phi)$ are the $z$-axisymmetric forms of the respective 3D expressions, which are now functions only of the polar angle $\phi$. If we assume that the surface of the particle can be partitioned into $i$ stripes, for which the effective molecule-surface interaction $\tilde{\Phi}^{\ast}_{k}(r,\phi)$ is piecewise constant, see Fig.~\ref{fig:bands}, then Eq.~\cref{eq:velocity} can be rewritten further. For species $k$, the $i$-th integration surface is denoted using $S_{i,k}$ and the relevant slip factor using $\xi_{k}(\phi_{i,k})$ which is given by Eq.~\cref{eq:couple} evaluated in a point in the region labeled by $\phi_{i,k}$. The velocity is then given by
\begin{eqnarray}
\nonumber v & = & -\sum_{k=2}^{N} \sum_{i} \xi_{k}(\phi_{i,k}) \\
\nonumber & & \phantom{ -\sum_{k=2}^{N} \sum_{i}} \times \iint_{S_{i,k}} \frac{ \left( \hat{\mathbf{t}}(\phi) \cdot \mathbf{\nabla} \omega_{k} (\phi) \right) \left( \hat{\mathbf{t}}(\phi) \cdot \hat{\mathbf{z}} \right) }{4 \pi a^{2}} \ud \mathbf{s}, \\
\label{eq:vel_part} & &
\end{eqnarray}
where the $\xi_{k}(\phi_{i,k})$ act as multiplicative constants on the respective domains of integration. This is a particularly useful form for the analysis of the effect of the molecule-surface interaction.

\section{\label{sec:perturb}The Slip Model and Perturbations of the Surface-Molecule Potential}

We conclude our discussion of the theory of self-diffusiophoretic drive by considering perturbations of the surface-molecule interaction potential for the slip model. A perturbation formalism can be used to quantify the effect of modifying the surface-molecule interaction in an experiment. For instance, by changing the material of the inert surface or adding surfactants to the system that adsorb to part of the SPP's surface the interaction potential is modified. If the impact of such a modification is small, the effect on the velocity of the colloid can be straightforwardly determined by utilizing the slip-layer solution for the unperturbed potential. Moreover, the perturbation theory can be used to quickly assess the change in speed and direction of an SPP by varying the surface-molecule interaction over a range of parameters. The relevance of this approach will become more clear in the results section. 

Let us assume that the interaction $\tilde{\Phi}_{1}(\phi)$ for the solvent remains unperturbed, i.e., only the interaction for the solutes may be modified. We further assume that $\xi_{k}(\phi)$ is piecewise constant in $\phi$ and that a solution to Eqs.~(\ref{eq:NS_sum}-\ref{eq:ma_sum}) with boundary conditions, Eqs.~(\ref{eq:unorm},\ref{eq:slip}), is given for a specific set of non-zero effective molecule-surface interaction potentials $\tilde{\Phi}^{\ast}_{k}(r,\phi_{i})$.

For small perturbations of the effective interaction potential $\tilde{\Phi}_{k}'(r,\phi_{i})$, where we keep the position of the piecewise constant domains fixed, the change with respect to the unperturbed effective potential $\tilde{\Phi}^{\ast}_{k}(r,\phi_{i})$ can be expressed using the effective coupling constant
\begin{equation}
\label{eq:coupeff} V_{\mathrm{eff},i,k} = \frac{\displaystyle \int_{0}^{\infty} r \left( \exp \left( -\tilde{\Phi}_{k}'(r,\phi_{i}) \right) - 1 \right) \ud r}{\displaystyle \int_{0}^{\infty} r \left( \exp \left( -\tilde{\Phi}^{\ast}_{k}(r,\phi_{i}) \right) - 1 \right) \ud r}.
\end{equation}
For small P\'{e}clet numbers one may safely assume that the effect of advective back coupling in Eq.~\cref{eq:cont_sum} is negligible, whereby small P\'{e}clet number we mean $Pe = a v / \tilde{D} \ll 1$ with $\tilde{D}$ as before. N.B. Colloidal SPPs are typically in the low P\'{e}clet number regime, as can be readily determined by examining the experimental parameters.~\cite{Brown,Ebbens,Ebbens_12,Howse_07,Valadares_10,simmchen} We refer to Ref.~\cite{Lauga_14} for a detailed discussion of the effects of a finite P{\'e}clet number on the self-propulsion. For $Pe \ll 1$, the distribution of species around the colloid does not change substantially by the introduction of the perturbation for the speed at which the particle is moving. This allows us to treat the
\begin{equation}
\label{eq:cons} C_{i,k} \equiv \iint_{S_{i,k}} \frac{ \left( \hat{\mathbf{t}}(\phi) \cdot \mathbf{\nabla} \omega_{k} (\phi) \right) \left( \hat{\mathbf{t}}(\phi) \cdot \hat{\mathbf{z}} \right) }{4 \pi a^{2}} \ud \mathbf{s}
\end{equation}
as constants with respect to the perturbation and write the self-propulsion velocity as a simple weighted sum over the interaction prefactors $\xi_{k}(\phi_{i})$ and species distribution terms
\begin{equation}
\label{eq:velred} v = - \sum_{i} \sum_{k=1}^{N} \xi_{k}(\phi_{i}) C_{i,k}.
\end{equation}
This leads to the following expression for the approximate self-propulsion velocity for the perturbed potentials in terms of the original solution
\begin{equation}
\label{eq:vapprox} v' = - \sum_{i} \sum_{k=1}^{N} V_{\mathrm{eff},i,k} \xi_{k}(\phi_{i}) C_{i,k},
\end{equation}
where for the $C_{i,k}$ we can use the values precomputed for the original potentials. Equation~\cref{eq:vapprox} is a powerful tool to evaluate the influence of perturbations in the molecule-surface interaction potential for small-P\'{e}clet-number SPPs, as we will see in the following.

\section{\label{sec:result}Results}

\subsection{\label{sub:paramet}System Parameters and Methods}

\begin{figure}[!htb]
\begin{center}
\includegraphics[scale=1.0]{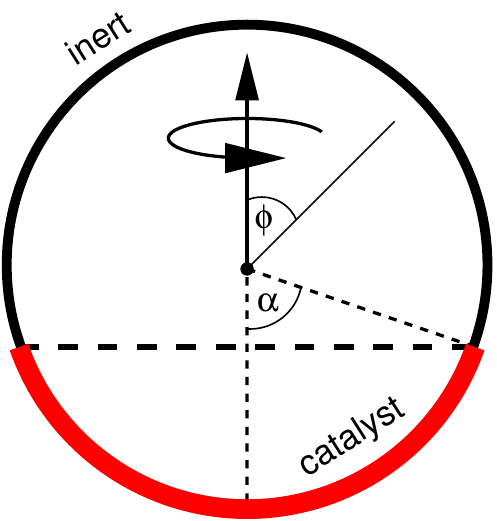}
\end{center}
\caption{\label{fig:alpha}(color online) Sketch of the level of catalytic surface coating. The figure is rotationally symmetric in the $z$-axis, as indicated by the arrow, and the catalytic surface (thick red line) coverage is given by the angle $\alpha$. The rest of the surface $\phi \in [0,\pi/2-\alpha]$ is assumed to be inert.}
\end{figure}

In order to study the effect of the molecule-surface interaction on the self-diffusiophoretic swimming speed of a colloidal SPP, we use the following quantities as a \emph{base} set. We consider $N = 3$ species for the multi-component model, with $i = 1$ (H$_{2}$O), $i = 2$ (H$_{2}$O$_{2}$), and $i = 3$ (O$_{2}$). The molecular mass of the components is given by $m_{1} = 18$ u, $m_{2} = 34$ u, and $m_{3} = 32$ u, respectively, with $\mathrm{u} = 1.66 \cdot 10^{-27}$ kg the atomic mass unit. The diffusion coefficients are given by $D_{1} = 2.3 \cdot 10^{-9}$ m$^{2}$\,s$^{-1}$~\cite{Holz}, $D_{2} = 1.2 \cdot 10^{-9}$ m$^{2}$\,s$^{-1}$~\cite{Stern,Schumb}, $D_{3} = 1.9 \cdot 10^{-9}$ m$^{2}$\,s$^{-1}$~\cite{Sridhar,St-Denis,Bartels}, respectively, where we used the values for (self-)diffusion in water. The colloid is assumed to be $a = 0.5 \cdot 10^{-6}$ m in radius. We set the density of hydrogen peroxide to 10\% b.v. A simple linear catalytic surface reaction is used as in Eq.~\cref{eq:reacbc}, with the stoichiometric and mass coefficients as in Eq.~\cref{eq:reacco}, i.e., 2H$_{2}$O$_{2}$ $\rightarrow$ 2 H$_{2}$O $+$ O$_{2}$, where the reaction rate is given by 
\begin{equation}
\label{eq:reaccoef} R(\phi) = 2.0\cdot10^{-5} \textrm{ } \mathrm{m\,s}^{-1} \times \left\{ \begin{array}{ccl} 0 & \, & \phi \le \pi - \alpha \\ 1 & \, & \phi > \pi - \alpha \\ \end{array} \right. ,
\end{equation}
with $\alpha$ the area of the colloid covered by the catalyst, see Fig.~\ref{fig:alpha}. The choice for this specific reaction rate is such that it reproduces the hydrogen-peroxide-depletion rates given in Ref.~\cite{Brown} for this H$_{2}$O$_{2}$ concentration; we obtain a consumption of $1.0\cdot10^{11}$ molecules/second per SPP. From the mass density of water $1.0 \cdot 10^{3}$ kg\,m$^{-3}$ and hydrogen peroxide $1.45 \cdot 10^{3}$ kg\,m$^{-3}$ at room temperature, it then follows that for this volume fraction of H$_{2}$O$_{2}$, the total density of the fluid can be approximated by $\rho = 1.045 \cdot 10^{3}$ kg\,m$^{-3}$. Similarly, we can use the formalism of Ref.~\cite{Bloomfield} to compute the dynamic viscosity of the mixture. Using the viscosity of water $1.0 \cdot 10^{-3}$ kg\,m$^{-1}$\,s$^{-1}$ and hydrogen peroxide $1.245 \cdot 10^{-3}$ kg\,m$^{-1}$\,s$^{-1}$ at room temperature, we arrive at $\eta = 1.017 \cdot 10^{-3}$ kg\,m$^{-1}$\,s$^{-1}$ for the mixture's dynamic viscosity. For these parameters we find that $\omega_{1,\mathrm{res}} = 0.861$, $\omega_{2,\mathrm{res}} = 0.139$, and $\omega_{3,\mathrm{res}} = 0.000$.

The above choices leave only four free parameters. The catalytic coverage of the colloid $\alpha$ and the three surface-molecule interaction potentials $\tilde{\Phi}_{k}(r)$. Both the full multi-component and the slip-layer model can be solved using, e.g., finite-element methods or kinetic lattice-Boltzmann schemes. In this manuscript, we solve our system of equations using the COMSOL Multiphysics 4.4 solver suite. For the slip model, we combine the slip boundary condition of Eqs.~(\ref{eq:unorm}-\ref{eq:couple}) with the full multi-component solution in the region where the surface-molecule interaction is zero. We refer to this as the hybrid slip-layer model. Finally, we consider the dilute-limit approximation, where we combine the slip model with Eqs.~(\ref{eq:unorm}-\ref{eq:couple}), but ignore the $m(\mathbf{r})$-related term in Eq.~\cref{eq:jast_sum}.

\subsection{\label{sub:compare}Comparison between the Full Multi-Component Model and the Dilute-Limit Slip Model}

Let us start by assuming that the interaction between the three species of molecules is uniform over the surface. We choose the following form for the surface interaction potential $\tilde{\Phi}_{3}(r)$ between the SPP's surface and the solvated oxygen
\begin{eqnarray}
\label{eq:f} f(r) & = & \frac{1}{r^{3}} - \frac{1}{(a+b)^{3}} ; \\
\label{eq:g} g(r) & = & f(r) - \frac{\partial}{\partial r}f(a+b) (r - (a+b)) ; \\
\label{eq:Phi3} \tilde{\Phi}_{3}(r) & = & \left\{ \begin{array}{ccl} 0 & \, & r < a \vee r \ge a + b \\ g(r)/g(a) & \, & a \le r < a + b \\ \end{array} \right.  ,
\end{eqnarray}
where $a$ is the colloid radius, set to 0.5 $\mu$m here, $b$ is 1 nm, and $f(r)$ and $g(r)$ are auxilliary functions. This corresponds to a repulsion that decays as $\tilde{\Phi}_{3}(z) \propto 1/z^{3}$, with $z$ the distance between the surface and the molecule, as before. The potential is set to 1 $k_{\mathrm{B}}T$ at contact ($z = 0$) and decays to $0$ over a length of 1 nm, in such a way that $(\partial/\partial z) \tilde{\Phi}_{3}(z) = 0$ at this distance; this decay is accomplished by construction via the auxilliary functions. We assume that the other interaction potentials are zero, i.e., the ones for the water and hydrogen peroxide. The particles thus only interact via soft short-ranged potentials, i.e., their interaction with the wall is that of point particles, which is modified by a slight repulsion in the case of the oxygen. The total force acting on the species, see Eq.~\cref{eq:ftotk_sum}, can now be straightforwardly calculated. 

Note that the above choice constitutes a rather simplified set of interaction potentials. Employing hard or strongly divergent potentials at the surface is not admissible in our model, since this would interfere with the way our incompressibility condition is constructed and the ideal gas behavior of the solutes. Any potential that attracts species sufficiently to the surface of the colloid would cause the solute to be strongly accumulated near the surface, since our theory does not incorporate an excluded-volume term. N.B., a strongly repulsive term would also cause problems, due to the back-coupling in the flux equations by the correction term, see Eq.~\cref{eq:ftotk_sum}. Moreover, employing such potentials would require a re-examination of the reaction scheme in the flux boundary conditions, since it would no longer be obvious at which distance the molecules can react with the surface. Therefore, it is ill-advised to incorporate anything but the soft-potential parts of the surface-molecule interactions, which is the reason behind the above choice.

It should also be emphasized that solving the full multi-component model for a system, where the interaction length is a factor of $10^{-3}$ smaller than the colloid diameter, is technically challenging. However, it is necessary in order to compare the validity of the slip-model results, as the slip-model model is only applicable when there is such a strong separation of length scales.~\cite{Mood_13} We therefore made a judicious choice in specifying the range of the interaction potential to be $1$ nm, rather than $\sim 2$ {\AA}, to keep the model computationally manageable. 

\begin{figure}[!htb]
\begin{center}
\includegraphics[scale=1.0]{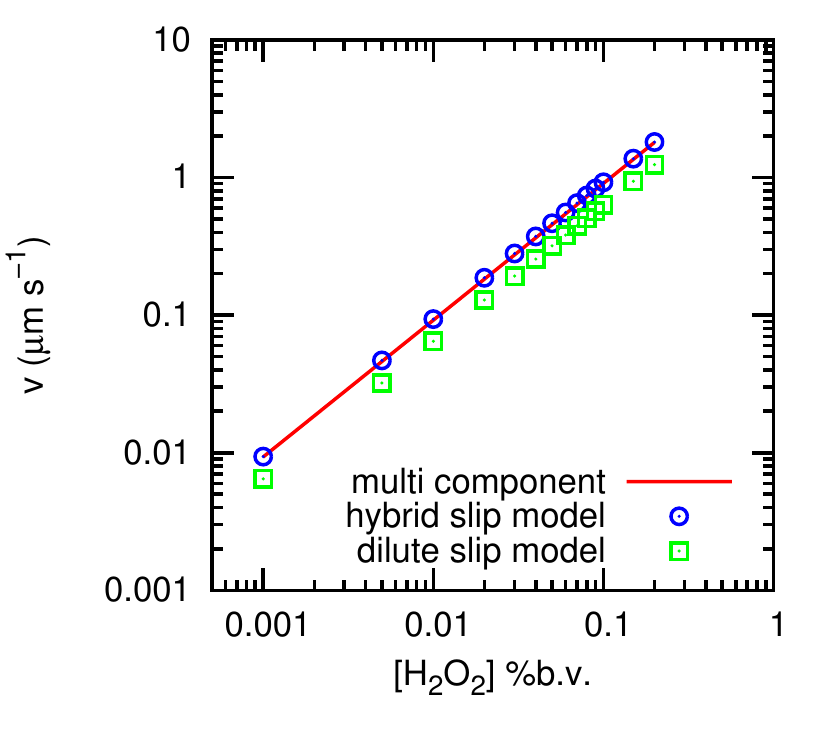}
\end{center}
\caption{\label{fig:versus}(color online) The dependence of the self-propulsion velocity $v$ on the imposed hydrogen peroxide concentration, expressed in a by-volume percentage. The red line shows the self-propulsion velocity that follows from solving the full multi-component model, the blue circles show the result of the hybrid slip-layer approximation, and the green squares show the result of the dilute slip-layer approximation.}
\end{figure}

We can assess the quality of the slip-model approximation, by solving the full set of multi-component equations and comparing this to the slip-model result for the same interaction parameters. In order to make the comparison, the requirements for the validity of the slip-layer approximation were checked and were found to be satisfied to within reasonable tolerance. In Fig.~\ref{fig:versus}, we consider the dependence of the terminal velocity on the concentration of hydrogen peroxide, while keeping all parameters the same for the three models (multi-component, hybrid, and dilute-limit). Note that the full multi-component model and our hybrid multi-component slip-layer approximation model match over the entire range of densities considered here. This is a surprising result, as it implies that use of the hybrid model is permissible for very high levels of solute concentrations. However, one should be careful not to generalize this result too far, as such correspondence may break down for other parameter sets or less well-behaved interaction potentials. Moreover, the use of too high concentrations will violate the approximations that went into our multi-component model.

Note that the dilute-limit approximation gives a qualitatively similar result, however, the velocity of the SPP is systematically underestimated by a factor of 30\%. This can be attributed to the absence of the cross coupling based on $m(\mathbf{r})$ in Eq.~\cref{eq:jast_sum}. Interestingly, the dilute limit result does not approach the solution of the full multi-component model in the limit of zero solute concentration. This implies that the cross-coupling term in Eq.~\cref{eq:jast_sum}  is \emph{not} sub-dominant even in the dilute limit and could significantly impact the self-propulsion velocity, when there are substantial differences in the mass of the molecular species. The level of correspondence is, however, good enough to justify the use of the equal-mass dilute-limit approximation for a hydrogen-peroxide solution.

For further discussion on the topic of multi-component modeling in the dilute limit, we refer the interested reader to the work by Sharifi-Mood~\emph{et al.}~\cite{Mood_13} In Ref.~\cite{Mood_13} the effect of the ratio between the interaction length and the colloid radius on the self-propulsion speed is examined and found to be significant when the radius of the colloid is sub-micron in size.

\subsection{\label{sub:single}Slip-Model Self-Propulsion for Hard Surface-Molecule Interactions}

From here on, we consider only the hybrid slip model. The reason for this is, that is it far less computationally expensive to derive results using this approximation and it therefore allows us to cover a larger region of parameter space. Moreover, we can employ less well-behaved interaction potentials, since the slip prefactors $\xi_{k}(\phi)$ can be precomputed analytically and numerical stability in the slip-layer is not a prerequisite. In fact, it is commonplace to relax the conditions on the potentials significantly, in order to permit hard-core interactions.~\cite{Brady_11} From a technical point, this is questionable, but we decided to follow suit here, in order to relate our result to previous studies. In this section, we still assume that the surface-molecule interaction is homogeneous. The molecule-surface interaction is of the form
\begin{equation}
\label{eq:poths} \tilde{\Phi}_{k}(r,\phi) = \left\{ \begin{array}{ccl} 0 & \, & r - a > a_{k} \\ \infty & \, & r - a \le a_{k} \\ \end{array} \right. ,
\end{equation}
where the $a_{k}$ is the molecular radius of the $k$-th component. It should be noted that Eq.~\cref{eq:phstdil} is only well-defined for hard interaction potentials, when we specify the way these can be subtracted, again being indicative of the inherent problems of using such potentials. In particular, to avoid infinities in $\xi_{k}(\phi)$, the solvent must have a smaller hard radius than all solutes, i.e., $a_{1} < a_{k>1}$. We then define the difference between the two infinities to be zero, such that the contribution of Eq.~\cref{eq:phstdil} to $\xi_{k}(\phi)$ is zero over the range $a < r < a + a_{1}$. Note that hard potentials can only be introduced in this manner into the slip-layer approximation, as the more complex expression for the total force of Eq.~\cref{eq:ftotk_sum} would result in an ill-posed problem.

\begin{figure*}[!htb]
\begin{center}
\includegraphics[scale=1.0]{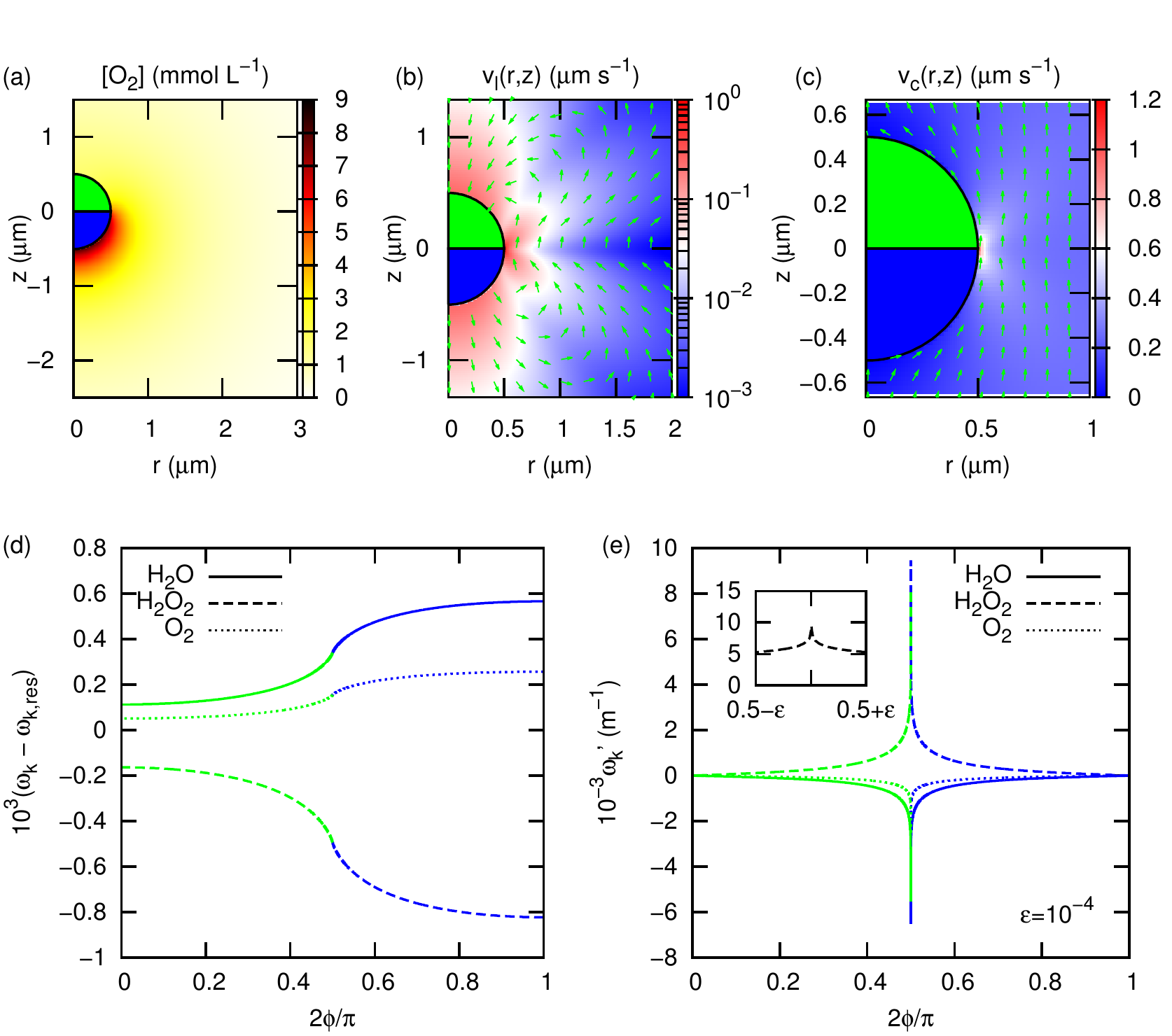}
\end{center}
\caption{\label{fig:hemi}(color online) Properties of the self-propulsion of a half-coated colloidal swimmer with radius $0.5$ $\mu$m, for the base set of parameters given in the text. (a) The oxygen concentration profile around the particle. The figure is rotationally symmetric in the $z$-axis. The bottom half (blue) of the particle is the catalytic side and the top half (green) the inert side. (b) The fluid velocity magnitude $v_{l}(r,z)$ in the lab frame. The flow field does not possess a monopolar contribution, since the self-propulsion is a force-free mechanism. It consists of higher-order hydrodynamic multipoles, with a far-field potential-swimmer nature. Green arrows show the direction of fluid flow. (c) Velocity profile in the co-moving frame $v_{c}(r,z)$. Note the increased fluid velocity close to the equator; the SPP is effectively moving forward. (d) The mass fraction $\omega_{k}$ of the three components along the surface of the SPP, as indicated using the polar angle $\phi$, relative to the mass fraction in the reservoir $\omega_{k,\mathrm{res}}$. The inert part $0 < \phi < \pi/2$ of the curve is color-coded green (left) and the catalytic part blue (right). (e) The mass-fraction gradient along the surface of the SPP, where we used the notation $\omega_{k}' \equiv \left(\hat{\mathbf{t}}(\phi) \cdot \mathbf{\nabla} \omega_{k} (\phi) \right)$; the color coding is the same as in (d). The inset shows a detail of the curve for hydrogen peroxide.}
\end{figure*}

In the following we typically use $a_{1} = 1.4$ {\AA}, $a_{2} = 1.9$ {\AA}, and $a_{3} = 1.8$ {\AA}.~\cite{israel} Figure~\ref{fig:hemi}a shows concentration profile of oxygen around the colloid for $\alpha = \pi/2$ and the above parameter choices. The maximum concentration of oxygen around the platinum cap is approximately $9$ mM, in agreement with the oxygen-production-rate measurements of Ref.~\cite{Brown}. The corresponding fluid velocity profile is shown in Fig.~\ref{fig:hemi}b, which gives $\mathbf{u}(\mathbf{r})$ in the laboratory frame, wherein the fluid is stationary and the particle moves. The velocity achieved for this system is $-0.25$ $\mu$m\,s$^{-1}$. That is, the particle is moving in the direction of the platinum cap with a speed of $0.25$ $\mu$m\,s$^{-1}$, in agreement with the estimate given in Ref.~\cite{Brown}. The calculated speed is substantially smaller than is typically observed in experiment for these parameters~\cite{Brown,Ebbens,Ebbens_12,Howse_07,Valadares_10,simmchen}, in particular Ref.~\cite{Brown} establishes the velocity to be around $11 \pm 6$ $\mu$m\,s$^{-1}$. Moreover, in the experiment the particle is moving in the direction of the uncoated surface. Unfortunately, it is difficult to extract information about the nature of the self-propulsion mechanism from the mismatch between this diffusiophoretic result and the experimental observations, as will become clear in the following.

The flow field in Fig.~\ref{fig:hemi}b displays the expected potential far-field contribution for force-free SPPs with finite extent.~\cite{Lauga_12} Here, the term potential signifies a dipole with finite extent. However, a combination of a dipolar and higher-order modes is found close to the surface. This result is similar to the thin-cap limit for self-thermophoresing particles studied in Refs.~\cite{Bickel,Yoshinaga}. We found that the potential-swimmer approximation only holds for distances $r > 10a$. Figure~\ref{fig:hemi}c shows the flow field in the co-moving frame. For the reasonable parameter choices made here, the dominant contribution to the self-propulsion velocity comes from a small region near the divide between the inert and catalytic material, which is only $\approx 50$ nm in size, see Fig.~\ref{fig:hemi}c. This can be attributed to the fact that the concentration gradient along the surface $\left( \hat{\mathbf{t}}(\phi) \cdot \mathbf{\nabla} \omega_{k} (\phi) \right)$ is the largest in this region, as is illustrated in Fig.~\ref{fig:hemi}d,e, which show the mass fraction and its gradient along the surface, respectively. 

Note that $\left( \hat{\mathbf{t}}(\phi) \cdot \mathbf{\nabla} \omega_{k} (\phi) \right)$ is extremely spiked in this region, as is further evidenced by the inset to Fig.~\ref{fig:hemi}e where a close-up of is shown. From our numerical results it is clear that this feature remains present at small length scales. However, it is not immediately evident whether this feature is a cusp or not, due to the complexity of the coupled system of equations that is solved simultaneously. Finally, it should be remarked that in applying the slip-layer approximation, the gradient is evaluated at a distance $\delta$ away from the surface, whereas Fig.~\ref{fig:hemi}e shows the result at the surface ($r=a$). Without advection, cross-coupling, and force-correction, the differential equation for the solute $\omega_{k}(\mathbf{r})$ is of the Harmonic form ($\mathbf{\nabla}^{2}\omega_{k}(\mathbf{r}) = 0$). For this equation the solution can be expressed in terms of a Legendre-polynomial expansion with radial dependence according to $(a/r)^{l+1}$, with $l\ge0$ the summation index. Therefore, the convergence of the expansion at $r = a + \delta$ is improved by the geometric factors $(a/(a+\delta))^{l+1} < 1$ for the problematic point $\phi = \pi/2$.

\begin{figure}[!htb]
\begin{center}
\includegraphics[scale=1.0]{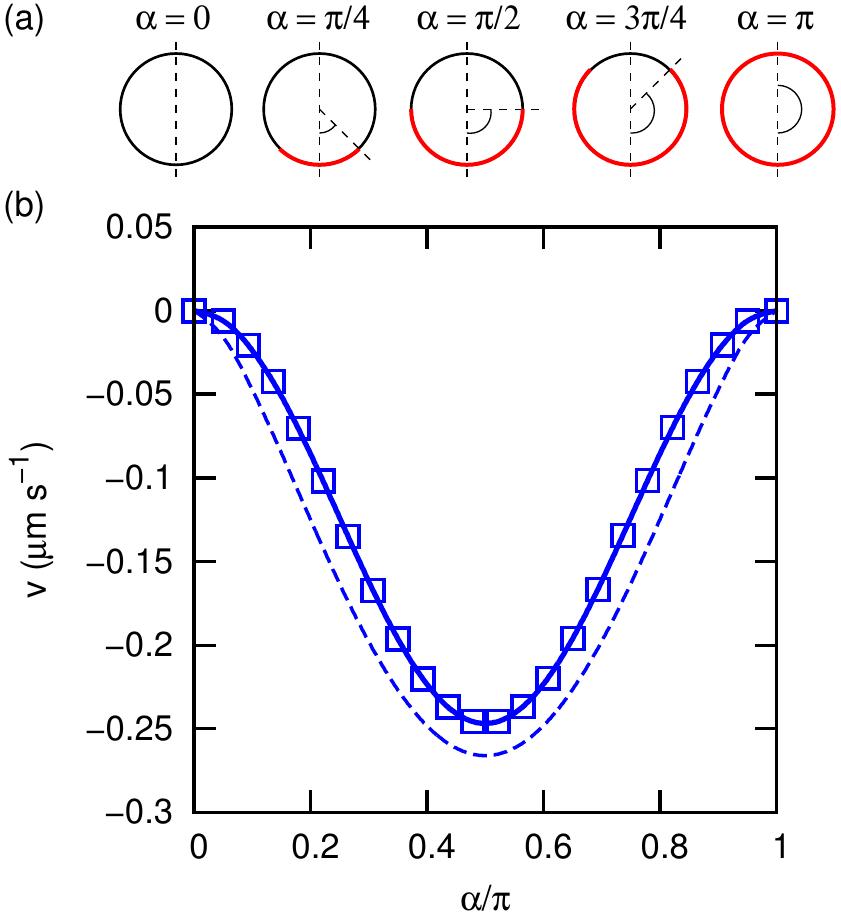}
\end{center}
\caption{\label{fig:base_alpha}(color online) Dependence of the self-propulsion velocity $v$ on the level of catalytic coating for our base parameter set. (a) Sketches of the level of catalytic coating, which is indicated by the labels and thick red curve. (b) The self-propulsion velocity of the colloid as a function of the coating parameter $\alpha$. The squares show the result of our simulations, the thick solid curve serves as a guide to the eye, and the dashed curve is the result of our low-P\'{e}clet-number approximation.}
\end{figure}

For our base choice of the interaction parameters, we find that there is a symmetry in the self-propulsion velocity as a function of the surface fraction of the colloid that is catalytically active (given by the angle $\alpha$), see Fig.~\ref{fig:base_alpha}b. Note that the estimate of our low-P\'{e}clet-number approximation is semi-quantitative. There seems to be a systematic overestimation of the magnitude of the velocity, but this is not a general property of the approximation, as can be seen in Fig.~\ref{fig:asym}b. The result in Fig.~\ref{fig:base_alpha}b is in accordance with literature results~\cite{Popescu_10,Mood_13}. We find a bell-shaped curve, in agreement with Ref.~\cite{Mood_13}, while in Ref.~\cite{Popescu_10} a parabolic curve was found. This can be attributed to a difference in the way the flux boundary conditions for the reaction are implemented in Ref.~\cite{Popescu_10}. Our bell-shaped result is robust with respect to changes in the surface-molecule interaction, provided this interaction potential is homogeneous over the surface. However, the assumption of homogeneity is unrealistic, as previously indicated, since in experiments the inert surface is typically composed of PMMA, silica, polystyrene, or SU-8, whereas the catalytic surface is composed of Pt. It is therefore reasonable to assume a different interaction between a molecule of species $k$ with either of the two surfaces. Unfortunately, it is unclear what the specific interaction is, since the full details of the molecule-surface interactions are typically not known. See, e.g., Refs.~\cite{Sabass_12a,Brown,Mood_13} for a discussion of the influence of interaction parameters for a half-coated SPP.

\subsection{\label{sub:hetero}Heterogeneity in the Surface-Molecule Interactions}

To address the problem of surface heterogeneity in the interaction potential, we take the following approach. Let us start by assessing the nature of a change in interaction potential, without introducing surface heterogeneity for now. We consider two molecule-surface interaction potential types: (i) a van der Waals (vdW) potential with a hard-core term $U_{\mathrm{vdW}}(s)$ and (ii) a simple square well $U_{\mathrm{sw}}(s)$, also with a hard-core contribution. The expressions for these potentials are as follows
\begin{eqnarray}
\label{eq:UvdW}  \tilde{\Phi}_{k}^{\mathrm{vdW}}(r,\phi) & = & \left\{ \begin{array}{ccl} h_{k} \left( \displaystyle \frac{a_{k}}{r-a} \right)^{3} & \, & r - a > a_{k} \\ \infty & \, & r - a \le a_{k} \\ \end{array} \right. ; \\
\label{eq:Usw}   \tilde{\Phi}_{k}^{\mathrm{sw}}(r,\phi) & = & \left\{ \begin{array}{ccl} 0 & \, & r - a > 2a_{k} \\ h_{k} & \, & 2a_{k} \ge r - a > a_{k} \\ \infty & \, & r - a \le a_{k} \\ \end{array} \right. ,
\end{eqnarray}
where $h_{k}$ is the contact value of surface-molecule interaction potential. The effective potential variants of Eqs.~(\ref{eq:UvdW},\ref{eq:Usw}) can be straightforwardly computed. The choice for a square-well potential that has width $a_{k}$ is admittedly arbitrary, but it will suffice for our purposes. To allow for both repulsive and attractive molecule-surface interactions, we let $h_{k}$ assume both positive and negative values, respectively. The cubic nature of the decay in the vdW-type potential is related to the shape of the vdW interaction between a point and a half space.~\cite{israel} The colloid can be approximated by a half space whenever the molecule is sufficiently close to the surface, due to the short-ranged nature of the vdW interactions.

\begin{figure}[!htb]
\begin{center}
\includegraphics[scale=1.0]{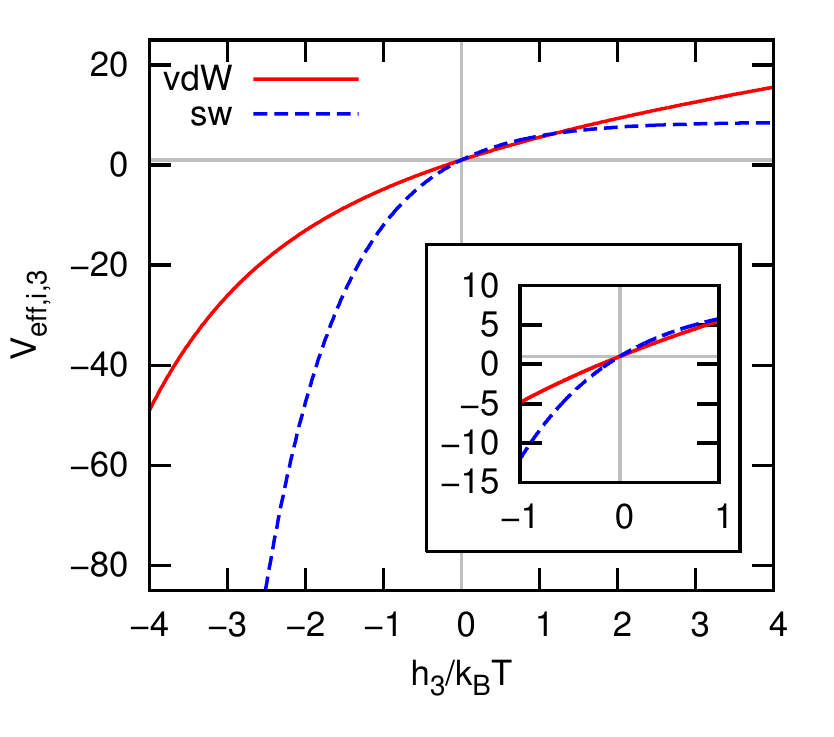}
\end{center}
\caption{\label{fig:c_cons}(color online) The effective coupling constant $V_{\mathrm{eff},i,3}$ for oxygen as a function of the contact value $h_{3}$ of the ranged soft interaction, by which the hard-core potential is modified. We considered two types of potential, a van der Waals-type interaction (vdW; solid red line) and a square-well (sw; dashed blue line) potential. The basis for comparison is the pure hard-core potential. The inset shows a close up for the range $h_{3} \in [-1,1]$. The solid gray lines indicate that the effective coupling constant is $1$ for $h_{3} = 0$.}
\end{figure}

We can now study the change in the self-propulsion velocity of the SPP as a function of the effective molecule-surface interaction potential. However, this still leaves for a large parameter space to be explored. We therefore make use of the concept of an effective coupling constant $V_{\mathrm{eff},i,k}$, as introduced in Eq.~\cref{eq:coupeff}, to reduce the number of degrees of freedom. Let us consider modifying the effective interaction of the oxygen, with respect to its hard-sphere value, i.e., $h_{3} = 0$ in Eqs.~(\ref{eq:UvdW},\ref{eq:Usw}). This hard potential will serve as our \emph{reference} potential. Then by varying $h_{3}'$ in Eqs.~(\ref{eq:UvdW},\ref{eq:Usw}), $V_{\mathrm{eff},i,3}$ can be easily calculated from Eq.~\cref{eq:coupeff}, see Fig.~\ref{fig:c_cons}. The first thing to note from the figure is the large spread in the effective coupling parameter, which assumes values in the range $[-60,15]$. In our calculation we assumed that the magnitude of the contact value of the interaction potential cannot exceed 4 $k_{\mathrm{B}}T$. This is a rather large value for simple non-charged molecule-surface interaction, however, even in the much smaller range of [-1,1] $k_{\mathrm{B}}T$, both the square-well and the vdW interaction show a tremendous variation of the effective coupling parameter, in the range $[-10,5]$. Qualitatively, adding a repulsive part to the potential $h_{k} > 0$ increases the effective size of the molecule, whereas adding an attractive part decreases it. When the attractive potential is sufficiently strong, the constant reverses sign. Recall that this effective coupling parameter acts as a multiplicative constant on the self-propulsion velocity achieved for the reference potential, see Eq.~\cref{eq:vapprox}, in the low-P\'{e}clet-number limit. A reversal of the sign could lead to the self-propulsion direction to invert. However, one must be careful, since the total velocity is the result of the interaction of three components with the surface of the SPP, for which the sign of the fuel concentration gradient is opposite to that of the products. 

\begin{figure*}[!htb]
\begin{center}
\includegraphics[scale=1.0]{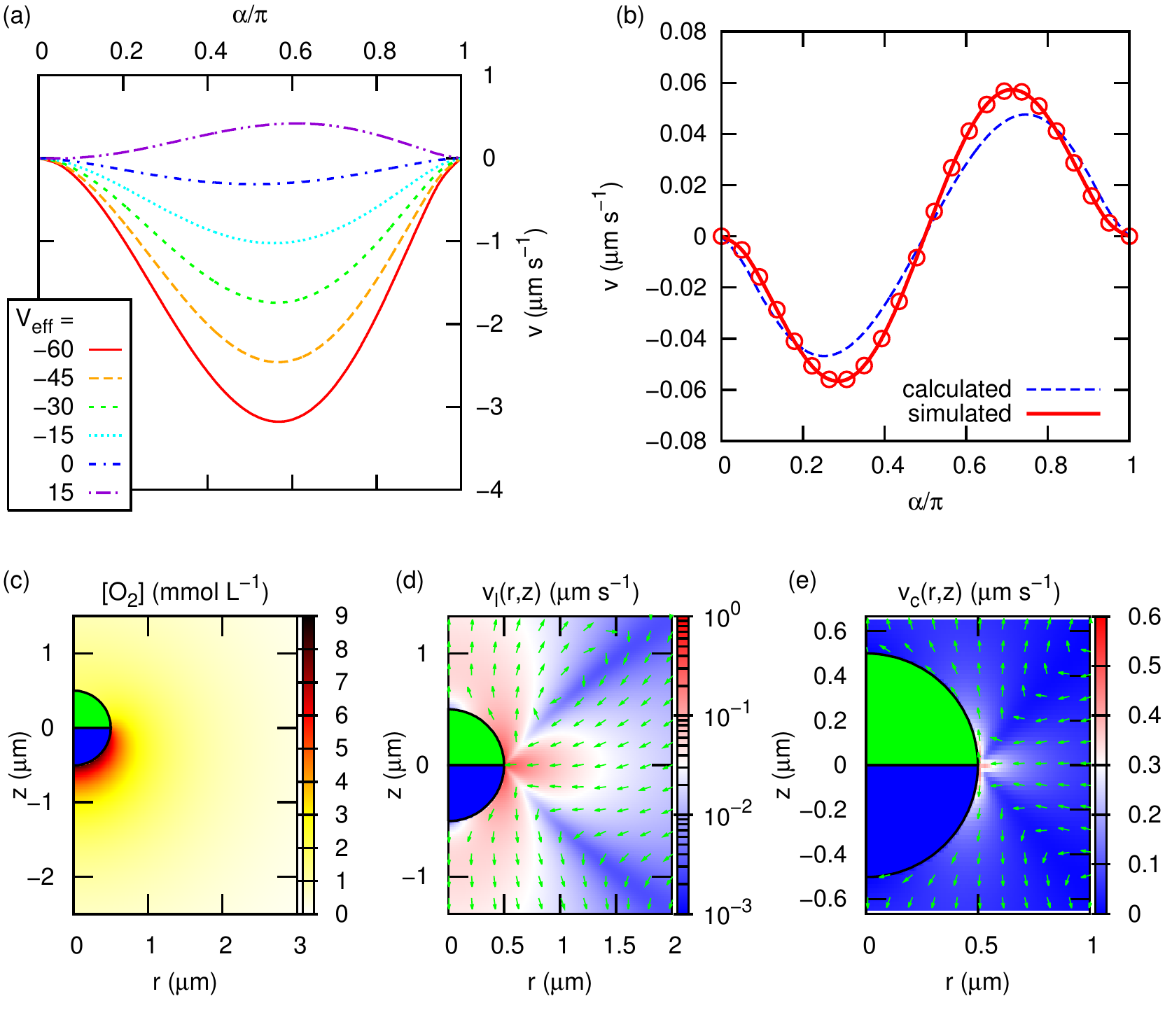}
\end{center}
\caption{\label{fig:asym}(color online) Dependence of the self-propulsion velocity on the level of catalytic coating for an asymmetric interaction potential, as given in the text. The interaction is given by the Eq.~\cref{eq:poths}, but the oxygen-platinum molecule-surface potential is modified according to Eqs.~(\ref{eq:UvdW},\ref{eq:Usw}). This is indicated using the effective coupling constant $V_{\mathrm{eff}}$. (a) The self-propulsion velocity $v$ as a function of the coating parameter $\alpha$ for several effective coupling constants. (b) The dependence of $v$ on $\alpha$ for $V_{\mathrm{eff}} = 6.8$. The red circles show the simulation data, the thick, red solid curve serves as a guide to the eye, the blue dashed curve follows from our low-P\'{e}clet-number approximation. (c) The oxygen concentration profile around the particle. The figure is rotationally symmetric in the $z$-axis, the blue half of the particle is the catalytic side and the green half the inert side. (d) The velocity profile $v_{l}(r,z)$ in the lab frame. Green arrows show the direction of fluid flow. Note that the flow field is almost purely dipolar. (e) Velocity profile in the co-moving frame $v_{c}(r,z)$, which is the same as the lab frame in this set up.}
\end{figure*}

Let us now consider the effect of introducing a small perturbation of the original molecular interaction potentials on the self-propulsion velocity as a function of the coating $\alpha$. We use $a_{1} = 1.4$ {\AA}, $a_{2} = 1.9$ {\AA}, and $a_{3} = 1.8$ {\AA}, with $h_{k} = 0$ for all surfaces, with \emph{one exception}. For the interaction between the oxygen and the catalytic side, we allow the radius and interaction strength to vary, such that $V_{\mathrm{eff},c,3}$, where the subscript $c$ indicates the catalytic region, assumes values between $-60$ and $15$, in agreement with the result of Fig.~\ref{fig:c_cons}. For this asymmetric interaction, the velocity profile is no longer symmetric with respect to $\alpha = \pi/2$, see Fig.~\ref{fig:asym}a. Moreover, we observe self-propulsion in the range between -3 $\mu$m\,s$^{-1}$ and 0.5 $\mu$m\,s$^{-1}$ depending on the amount of repulsion and attraction between the oxygen and the catalyst. That is, motion both in the direction of the platinum cap and in the direction of the inert part is observed. We investigated other combinations of molecule-surface interactions and found the trend of a large impact on the magnitude and direction of the swimming to be preserved.

A particularly unusual form of $\alpha$-dependence is found for $V_{\mathrm{eff},c,3} = 6.8$, see Fig.~\ref{fig:asym}b. This effective coupling corresponds to $h_{3} = -1.32$ $k_{\mathrm{B}}T$ and $h_{3} = -1.45$ $k_{\mathrm{B}}T$ in the vdW and square-well models, respectively, for $a_{3} = 1.8$ {\AA}. For $V_{\mathrm{eff},c,3} = 6.8$ the self-propulsion velocity of the half-coated Janus colloid, is not just sub-optimal, it is zero. The maximum propulsion speed is 0.06 $\mu$m\,s$^{-1}$, which is in the direction of the platinum cap for a coating of $\alpha \approx 0.3 \pi$ and in the direction of the inert part for $\alpha \approx 0.7 \pi$. This implies that within the confines of the presented self-diffusiophoretic propulsion model, highly unexpected dependence of the SPP's velocity can be encountered using only small perturbations of the molecule-surface potential. Note that the predictions of our simple scaling argument, see Eq.~\cref{eq:vapprox}, and the full slip-model calculations is quite excellent in Fig.~\ref{fig:asym}b. This shows the strength of the method as a means to investigate the self-propulsion properties of a low-P\'{e}clet-number system as a function of the interaction potential. For completeness, we have added the oxygen concentration profile and the fluid velocity profile in the lab and co-moving frame for the half-coated Janus particle, $\alpha = 0.5 \pi$, in Figs.~\ref{fig:asym}c,d,e, respectively. It is immediately apparent that the potential-swimmer asymmetry in the flow field for the particle of Fig.~\ref{fig:hemi}c is no longer present. The particle has an almost dipolar velocity field and acts as a `shaker', a particle that does not move but causes fluid flow around itself. 

\subsection{\label{sub:reaction}Sensitivity of the Self-Propulsion Speed to the Particulars of the Catalyst's Reactivity}

\begin{figure}[!htb]
\begin{center}
\includegraphics[scale=1.0]{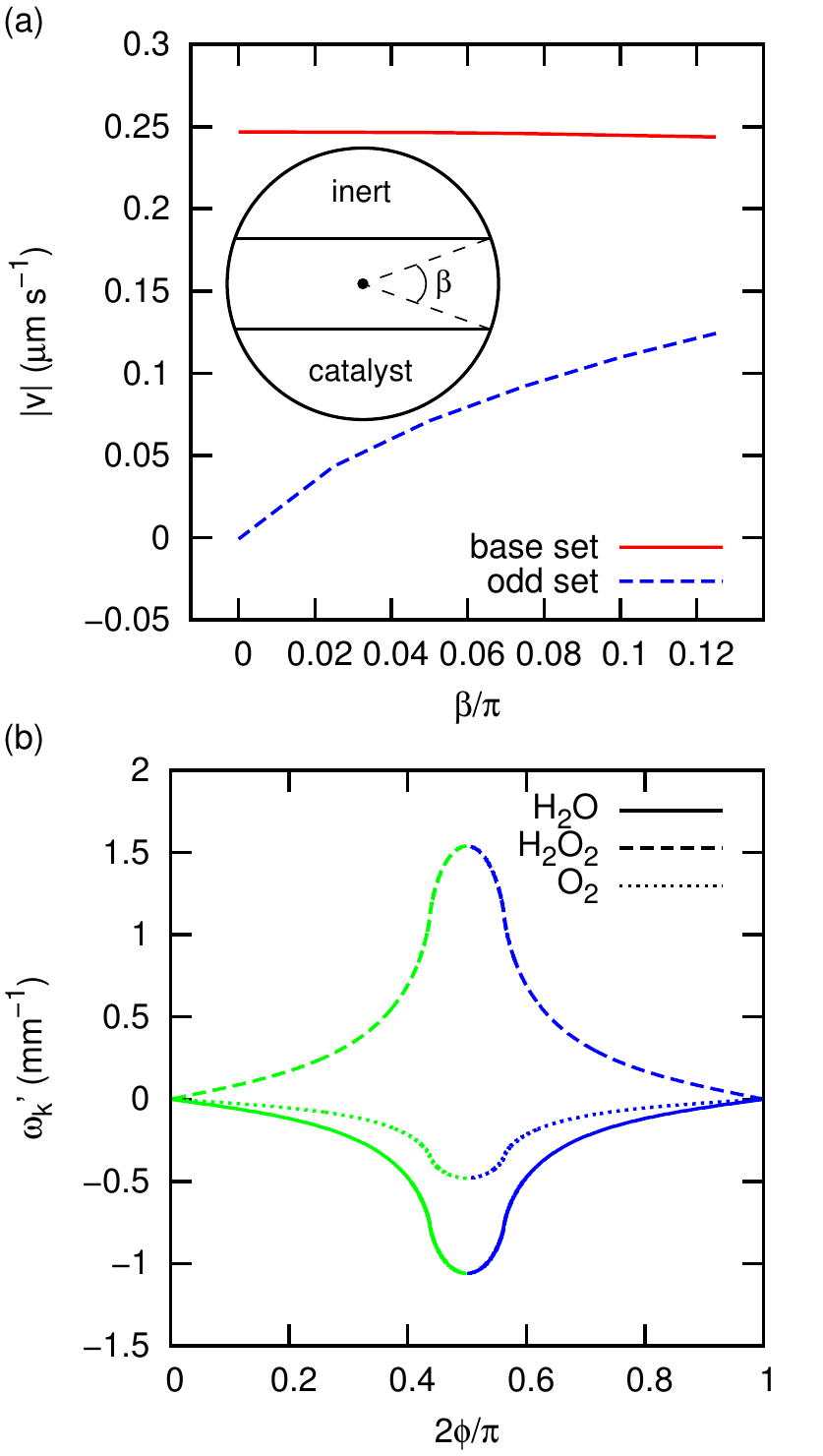}
\end{center}
\caption{\label{fig:beta}(color online) (a) The dependence of the self-propulsion velocity $v$ on the smoothing parameter $\beta$. The sketch in the inset shows the range over which the catalytic reaction rate linearly changes from its value on the catalytic side to zero on the inert side. The solid red curve gives the dependence for the base set, while the blue dashed curve shows the dependence for the system of Figs.~\ref{fig:asym}c-e. (b) The mass-fraction gradient along the surface of the SPP for $\beta/\pi = 0.12$, where we used the notation $\omega_{k}' \equiv \left(\hat{\mathbf{t}}(\phi) \cdot \mathbf{\nabla} \omega_{k} (\phi) \right)$. There is little difference between the two data sets in this curve, therefore we only show the one corresponding to the base set. The color coding is the same as in Fig.~\ref{fig:hemi}(d,e).}
\end{figure}

Finally, let us examine the consequence of the localized nature of the slip contribution to the particle speed. That is, how sensitive is the self-propulsion velocity to small variations of the reactivity around the part where the catalytic cap meets the inert surface of the colloid. We assume the base parameter set with the homogeneous hard molecule-surface interaction of Eq.~\cref{eq:poths} and molecular radii $a_{1} = 1.4$ {\AA}, $a_{2} = 1.9$ {\AA}, and $a_{3} = 1.8$ {\AA}, as before. However, we now assume the following form for the reaction rate
\begin{eqnarray}
\nonumber R(\phi) & = & 2.0\cdot10^{-5} \textrm{ } \mathrm{m\,s}^{-1} \times \\
\label{eq:reaccoef_new} & &  \left\{ \begin{array}{ccl} 0 & \, & \phi \ge \pi/2 + \beta/2 \\[0.5em] \left( \displaystyle \frac{\pi + \beta - \phi}{2\beta} \right) & \, & \begin{array}{c} \pi/2 + \beta/2 > \phi \\ \phi\ge \pi/2 - \beta/2 \end{array} \\[1.5em] 1 & \, & \phi < \pi/2 - \beta/2 \\ \end{array} \right. ,
\end{eqnarray}
where $\beta$ is a small angle around the equator of the Janus colloid, see the inset to Fig.~\ref{fig:beta}(a), for which the reaction rate increases linearly along the surface from zero to the full rate. This rate dependence can be thought of as a rough model for the effect of the thinning in the Pt-coating around the equator due to the vapor-deposition procedure by which these particles are fabricated.~\cite{Ebbens} As can be seen from Fig.~\ref{fig:beta}(a), in which the $\beta$ dependence is given, the self-propulsion speed and direction are remarkably robust to local changes in the reactivity of the catalytic surface in the region where the dominant contribution to the slip velocity is found. This implies that the above sensitivity to the interaction potential, is \emph{not} a consequence of the sharp transition in the reactivity of the surface. 

Naturally, species diffusion smoothes out the step-like nature of the reaction boundary condition. However, this did not imply that the gradient of the concentration along the surface was free of strongly spiked gradients, as is shown in Fig.~\ref{fig:hemi}e and its inset. For the gradual change in reactivity of Eq.~\cref{eq:reaccoef_new}, the $\left( \hat{\mathbf{t}}(\phi) \cdot \mathbf{\nabla} \omega_{k} (\phi) \right)$ profile is less spiked and even rounded at the tip, see Fig.~\ref{fig:beta}(b). The lack of substantial change in the velocity by smoothing $\left( \hat{\mathbf{t}}(\phi) \cdot \mathbf{\nabla} \omega_{k} (\phi) \right)$ is indicates that the localized high value of the concentration gradient is relevant, but not its exact shape. This is a useful result, as it implies that slight local modification of the transition between the catalytic and inert surfaces can be employed to make the system more tractable numerically. Moreover, it shows that while the sharp feature in Fig.~\ref{fig:hemi}e may be unphysical, a result that is smooth on physical length scales can be readily obtained. 

To conclude our examination of the influence of the reactivity, we performed the same analysis for the $V_{\mathrm{eff},c,3} = 6.8$ system and found a substantially stronger dependence on $\beta$, see Fig.~\ref{fig:beta}(a). This was to be expected, since the unusual $\alpha$ dependence and close-to-zero velocity for $\alpha = \pi/2$ can only be achieved by careful tuning of the interplay between concentration gradient and interaction potential. Therefore, some care must be taken in applying a local smoothing of the inert-catalyst transition.

\section{\label{sec:disc}Discussion and Outlook}

Summarizing, in this manuscript we introduced a multi-component description for the diffusiophoretic self-propulsion mechanism of colloidal particles. We started by deriving a multi-component model for bulk fluids. The model incorporates hydrodynamic flow through the incompressible Navier-Stokes equations, which are coupled to advection-diffusion equations for the individual components. The advection-diffusion formalism was modified to ensure incompressibility of the total fluid, in a manner similar to that introduced in Ref.~\cite{Bird}. To this bulk model we added boundary conditions for diffusiophoretic motion of arbitrary shapes. The use of these boundary conditions was illustrated for the specific case of a partially Pt-coated colloidal Janus sphere suspended in hydrogen-peroxide solution. Following this, we considered the dilute limit for the solute concentration and derived the slip-layer model.~\cite{Anderson,Golestanian_07,Brady_08,Popescu_10,Brady_11,Ebbens_12,Mood_13,Brady_13,Brady_14,Lauga_14} Here, we showed that our derivation yields a different effective interaction potential than was found in Ref.~\cite{Mood_13}. We consider our effective interaction potential to be the correct one, by way of our derivation. Finally, we introduced a low-P{\'e}clet number perturbation theory for the self-diffusiophoretic velocity of a spherical colloid, for small modifications of the surface-molecule interaction potential.

Using the models that we introduced, we examined the self-propulsion velocity of a spherical SPP for a set of reasonable physical parameters. We started by considering the differences between the full multi-component model, the hybrid slip-layer model, and the dilute-limit slip-layer approximation, for high concentrations of hydrogen peroxide. For simple, well-behaved potentials we showed that there is excellent correspondence between the full multi-component model and the hybrid slip-layer model. In fact, the correspondence holds over a surprisingly large range of reactant concentrations. Moreover, the dilute-limit approximation shows qualitative agreement with the hybrid and full multi-component models. This indicates that previously used dilute-limit analysis of the behavior of self-diffusiophoretic colloids is probably more reasonable for high solute concentrations than one would expect it to be for the hydrogen-peroxide system. However, for situations were there is a larger discrepancy between the molecular masses of the species that make up the total fluid, the correspondence may be lost, as the mass cross-coupling term in the flux contributes for all densities. Moreover, it is known that there can be strong differences between the results of the slip-layer approximation and fully resolved multi-component models for small colloids ($a <$ 100 nm).~\cite{Mood_13} Therefore, some care must be taken not to extrapolate our result too far. Further examination of the quality of the slip-layer approximation for more complex systems is left for future study.

Next, we considered the dependence of the self-propulsion velocity on the level of catalytic coating and the molecule-surface interaction potential of the solvent and solutes using the hybrid slip-layer model. The choice of using this approximation is based on the speed by which one can examine parameter space to obtain a qualitative picture of the nature of the SPP's behavior. 

We obtained a similar dependence of the velocity on the level of catalytic coating as was found in Refs.~\cite{Popescu_10,Mood_13} for a homogeneous molecule-surface interaction. That is to say, the molecules interact in the same way with the inert part of the SPP as with the catalytic part. The deviations between our results and those of Ref.~\cite{Popescu_10} are reasonable considering the differences in the approach to modeling the self-propulsion mechanism. We obtained a perturbed potential-swimmer fluid velocity profile around the SPP, which is in accordance with literature results for certain types of thermophoresing SPPs.~\cite{Bickel,Yoshinaga} It should be noted that the far-field potential (extended dipole) flow field is only achieved for distances to the SPP exceeding 5 times the particle diameter. We also found that for small changes in the surface-molecule interaction potential, sizable changes in the self-propulsion velocity of the active particle can be achieved. Not only can the particle speed up, but it can also change the direction in which it is moving.

We extended our investigation into the effect of the catalytic surface coating on the self-propulsion velocity, by including anisotropy of the molecule-surface interaction. A strong asymmetry is found in the velocity of the SPP as a function of the surface coating, when the solute or solvent species interact differently with the catalytic and inert surface. Such a variation in interaction can be expected on physical grounds, because of the difference in material properties between the Pt-surface (catalyst) and the inert surface -- which typically consists of SiO$_{2}$, polystyrene, PMMA, or SU-8. To assess the nature of the interaction between the solvent and solute molecules, we considered an effective coupling parameter for the slip model. We determined sensible parameters for the coupling, by considering a hard-core interaction plus a short-ranged van der Waals and square-well type attraction/repulsion, respectively. For reasonable values of the interaction asymmetry, the asymmetry of the velocity in the surface coating can be quite pronounced and even fully asymmetric. In the latter case, the SPP moves towards the inert side when it is less than half coated, it moves towards the platinum side when it is more than half coated, and is immobile when it is half coated. The half-coated particle acts as a `shaker', i.e., it causes fluid flow but does not move. Interestingly, the shaker has a close-to-dipolar flow field. 

Our results have strong implications for the preparation and analysis of experimental systems, since there will be a natural difference in the molecule-surface interaction between the inert and catalytic parts of the surface. In an experiment where the level of the particle's catalytic surface coating can be varied, it should be expected that an asymmetry in the self-propulsion is observed. However, it could be possible to exploit the strong dependence of the self-propulsion velocity on the molecule-surface interaction by modifying the properties of the inert surface chemically. The addition of ligands that preferentially adsorb to the inert surface could, for instance, be applied to change the speed and possibly the direction of motion of the particle, since only small changes in the interaction potential are necessary to accomplish this. Furthermore, it might be possible to cause a particle to move forward and backward between two regions, by reversible modification of the surface in this manner. 

It should also be noted that the sensitive nature of the velocity on the molecule-surface interaction makes it difficult to assess the nature of the self-propulsion mechanism from experiments. The fact that the self-diffusiophoretic model does not predict velocities that are close to the ones observed in experiment for Pt-coated particles in hydrogen-peroxide solution and reasonable interaction parameters, is a strong indication that other mechanisms must be considered.~\cite{Brown,Ebbens} However, changing the interaction between two or more of the surfaces and the solvent and solute species, could lead to greater speed-ups for motion in the direction of the platinum coated side than we observed in this work. In particular, the addition of further (possibly charged) components to the system will frustrate the interpretation of the results. 

In conclusion, our results show the sensitive nature of the self-propulsion velocity on the molecular details in the self-diffusiophoretic model for the motion of a catalytically coated sphere, which may have interesting consequences for experimental systems. Future studies will focus on extending these findings to anisotropic particles, such as cone- and stomatocyte-shaped motors, as well as to the interaction between several SPPs that are driven by (ionic) self-diffusiophoresis and self-electrophoresis. 

\section*{\label{sec:ack}Acknowledgements}

J.d.G. acknowledges financial support by a ``Nederlandse Organisatie voor Wetenschappelijk Onderzoek'' (NWO) Rubicon Grant (\#680501210). We thank the ``Deutsche Forschungsgemeinschaft'' (DFG) for financial funding through the SPP 1726 ``Microswimmers -- from single particle motion to collective behavior''. We would also like to thank M. Popescu, A. Brown, and M. Cates for useful input concerning the multi-component modeling.

\end{document}